# Graph neural network explanations reveal a topological signature of disease-associated hubs in biological networks

**Kyle Higgins[1], Ivan Laponogov[1], Dennis Veselkov[1,2], Kirill Veselkov[1,3*]**

[1]Division of Cancer, Department of Surgery and Cancer, Faculty of Medicine, Imperial College London, London, UK.
[2]Department of Computing, Imperial College London, U.K.
[3]Department of Environmental Health Sciences, Yale University, New Haven, CT, USA.
[*]To whom correspondence should be addressed: kirill.veselkov04@imperial.ac.uk

**Abstract**

Graph neural networks (GNNs) are increasingly used to model biological systems, yet the reliability of post-hoc explanation methods for recovering meaningful molecular mechanisms remains unclear. Here, we systematically evaluate four widely used approaches: Saliency Attribution (SA), Integrated Gradients (IG), GNNExplainer, and Layer-wise Relevance Propagation (LRP) in identifying disease-relevant structure in breast cancer RNA-seq data projected onto a protein–protein interaction network. Using synthetic benchmarks with known ground-truth motifs, we show that explanation methods recover distinct signal organizations: SA performs best for sparse single-node drivers, whereas IG and LRP preferentially recover distributed pathway-like and cascade-like signals. In TCGA BRCA data, we identify a consistent topological signature of disease-associated hubs in which attribution peaks in the immediate 1-hop neighborhood and decays across successive network shells, a pattern most pronounced for IG and LRP and associated with strong enrichment of known cancer hubs. We further observe a trade-off between local hub enrichment and global gene ranking performance, with IG optimizing local enrichment and SA achieving superior global discrimination. Motivated by these complementary behaviors, we introduce a framework combining a shell-based hub score with consensus ranking across explainers. Consensus scores improve prioritization of canonical cancer genes (TP53, BRCA1, ESR1, MYC), reduce dependence on node degree, and, especially when tuned, outperform individual methods, while pathway enrichment reveals improved recovery of biologically coherent cancer programs, including ERBB2, RTK, and MAPK signaling along with immune and cytokine signaling, indicating tumor microenvironment and inflammation. Together, these results demonstrate that topology-aware integration of graph explanations can improve biological interpretability and biologically relevant molecular recovery.

**Background**

The molecular interactions that drive disease do not act as the sum of isolated effects but propagate through structured biological systems, including protein-protein interaction networks (PPI).[1,2] These networks are known to have modules (discrete bundles of components for a distinct functional task, often overlapping)[3,4], hierarchical organization[5-7], and scale-free topology[5,8], where networks are thought to expand continuously by the addition of new nodes and new nodes attach preferentially to already well-connected nodes, forming hubs.[9] This results in a network which, compared to random networks, is highly tolerant to random attacks but extremely sensitive to targeted attacks of important network components, including hubs.[8,10] Such attacks are the hallmarks of many diseases such as cancer, resulting in catastrophic network failure when present.[2,11] Therefore, identifying the hallmarks of network disruption in disease is of the utmost importance to determine disease mechanism.

Graph neural networks provide a natural framework for integrating omics measurements with prior biological network structure. [12-16] In the case of PPIs omics data such as RNA-seq, methylation, copy number variant (CNV), polymorphism, and other data can be projected onto existing PPI databases such as the Search Tool for the Retrieval of Interacting Genes (STRING)[17,18] database.[19-21] This representation allows molecular variation to be interpreted not only at the level of individual genes, but also in relation to their network neighbors and pathway context. By propagating features across PPI or functional association graphs, GNNs can capture nonlinear relationships between local gene-level changes and broader biological programs. This is particularly relevant in cancer, where tumor phenotypes often reflect coordinated disruption of pathways, signaling neighborhoods and regulatory modules.

Interpreting graph neural networks remains a challenge in this setting because predictions arise from both node (and potentially edge) features as well as graph structure, after multiple rounds of message passing and nonlinear transformation. By default, most architectures (excepting attention transformers) are inherently 'black boxes', a variety of explainer tools have been developed by the graph AI community which have promise in omics applications. Several methods are frequently used but vary depending on their mathematical formulations and emphasis on different graph behavior. Some of the most frequently implemented methods include saliency attribution (SA)[22,23], Integrated Gradients (IG)[24], GNNExplainer[25], and Layer-Wise Relevance Propagation (LRP)[26,27].

SA is one of the simplest gradient-based explanation methods. Implementing a partial derivative of the model prediction with respect to an individual feature (therefore gradient across the whole), it measures how strongly the model output changes with an infinitesimal change to input. In a graph classification setting, this results in a node-level feature map of local sensitivity. Due to the measurement of local sensitivity, it is particularly suited to identifying sharp signals driven by a single highly influential node or feature of a node.[28] However, due to the purely local derivative, it can overlook more distributed evidence and be unstable in nonlinear graphs. In a message-passing setting, it may overemphasize points where the model is most sensitive rather than the full set of nodes contributing to the sensitivity.

IG expands on the basic premise of SA by accumulating gradients along a path from baseline input to the observed samples. Instead of only relying on the derivative at the final point, it integrates over many intermediate inputs, giving a more global attribution than SA while remaining gradient-based.[28] This improves the detection of relevant signals in nonlinear models, especially where saturation effects are present. IG is therefore well-suited for datasets where the coordinated contribution of several nodes contributes to the prediction.

GNNExplainer takes a conceptually different approach compared to both SA and IG. Rather than computing gradients by some method with respect to the input alone, it learns a minimum mask over nodes, edges, or features that preserves the models prediction (insofar as this is possible). Its objective is framed in terms of maximizing mutual information (MI) between model and masked subgraph or features.[25] That is to say this method searches for a sufficient explanation (subgraph or features) of the prediction given.

LRP takes a fundamentally different perspective to all three. LRP starts from the model input and redistributes the prediction back through the network, layer-by-layer, adopting a conservation rule that relevance at each layer must sum to the total relevance at the previous.[26] The motivation is to track how information flowed through the actual computations performed in the network. The result is an attribution map with final prediction decomposed across input nodes and features (and optionally edges). This makes the design particularly suitable for capturing complex and subtle interactions between diffuse signals that propagate into significant network effects after multiple convolutions.

Despite growing use of GNNs in network-based omics, it remains unclear whether post-hoc explainers recover meaningful biological structures or mainly reflect explainer-specific bias, graph topology or model artefacts. To address this gap, we compared SA, IG, GNNExplainer, and LRP on synthetic graph motifs with known ground truth: sparse drivers, hubs, pathways and cascades. We then applied the same explainers to a TCGA breast cancer RNA-seq GNN projected onto a protein association network, asking whether attributions localize around canonical cancer genes. Finally, we developed a shell-based hub score and consensus ranking approach to prioritize candidate disease-associated hubs. This framework separates performance on controlled benchmarks from biological plausibility in real data, and treats resulting hubs as hypotheses for further validation.

## Results

### *Evaluating Explainers on Synthetic Data*

To systematically assess the performance of explainability methods in the presence of biological motifs, we evaluated performance across four signal archetypes: needle (‘in-the-haystack’), hub, pathway, and cascade. They represent increasing levels of biological complexity, with needle confining signal to a single node, hubs representing a single node which drives the expression of its immediate neighbors (like a transcription factor), pathways representing a generic subsection of the graph which is upregulated as a whole, and cascades representing a hierarchical mechanism with increasing expression from root to leaf. In all settings, an underlying GNN achieved perfect test accuracy and balanced accuracy on their respective synthetic tasks, indicating all downstream differences in explainer performance reflect explanation behavior rather than predictive failure. (For a full summary of results, see **Supplementary Table ST1.**)

### *Synthetic ‘Needle-in-the-Haystack’*

In the ‘needle’ setting, we evaluate the utility of all methods at detecting a single driver gene. (**Figure 1A**) All signal is confined to a single node, meaning that the classification label switches from 1 to 0 depending on whether an example node (node 3) has expression above or below 0. In this case, all methods achieve a top-5 absolute hit rate of 1.0, demonstrating they all work well for this simple task, with a top-1 hit rate of 0.7 for LRP and IG, 0.9 for GNNExplainer, and 1.0 for SA, demonstrating that SA performs the best overall for detecting single-node drivers of signal. These results confirm that all methods can recover trivial signals, but there are differences present in their ranking of the signals.

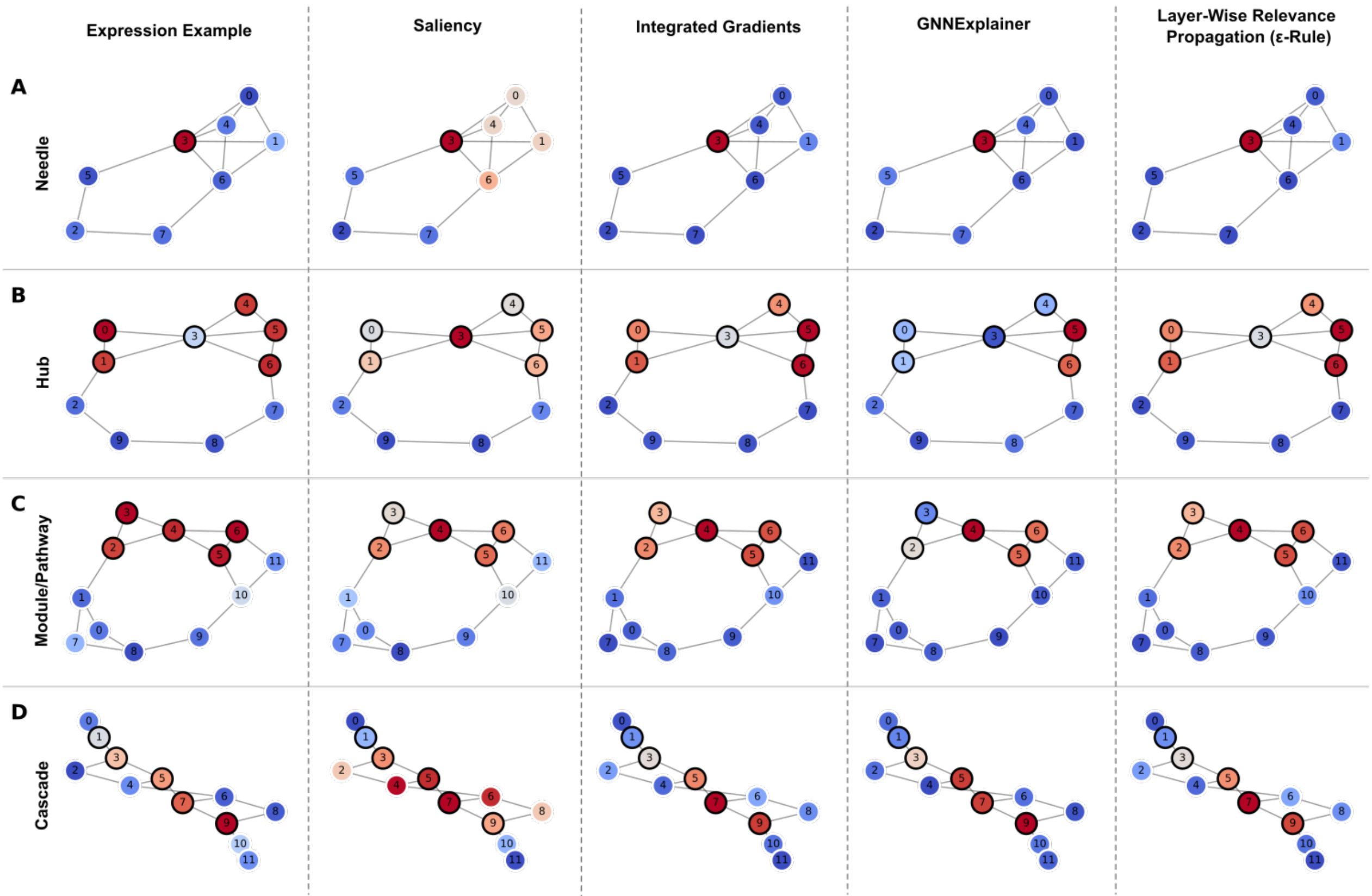

**Figure 1. Representative explanations across controlled biological signal archetypes.** Node-level attribution maps for four synthetic graph signal types: (**A**) needle, (**B**) hub, (**C**) pathway, and (**D**) cascade. Columns correspond to input feature values and explanation methods (SA, IG, GNNExplainer, and LRP-ε). Node color intensity reflects attribution magnitude (red, high importance; blue, low), and ground-truth signal nodes are outlined in black. Distinct attribution patterns are observed across methods, particularly for distributed and structured signals (pathway and cascade), where IG and LRP produce more spatially coherent attribution profiles compared to SA and GNNExplainer.

*Synthetic Hubs*

In the hub setting, where signal is concentrated in a highly connected node and its local neighborhood, we express this central node moderately highly with higher expression in its 1-hop neighborhood (representing the usually lower expression of signaling molecules like TFs). (**Figure 1B**) Here, differences become more pronounced, with mean absolute fraction on true nodes showing 0.92 for LRP and 0.93 for IG and mean positive fractions approaching 0.98, demonstrating a strong recovery of the full hub structure. (**Figure 2**) We observe 0.77 for SA, showing a decent capture of hub structure, however the method does well to recover the central node. Finally, we observe 0.68 for GNNExplainer despite perfect hit rates, indicating it often diffuses signal to other adjacent nodes. These results suggest that IG and LRP may be better suited to capture distributed topological influence distributed across a neighborhood, while SA is suited to identify the central node with high importance.

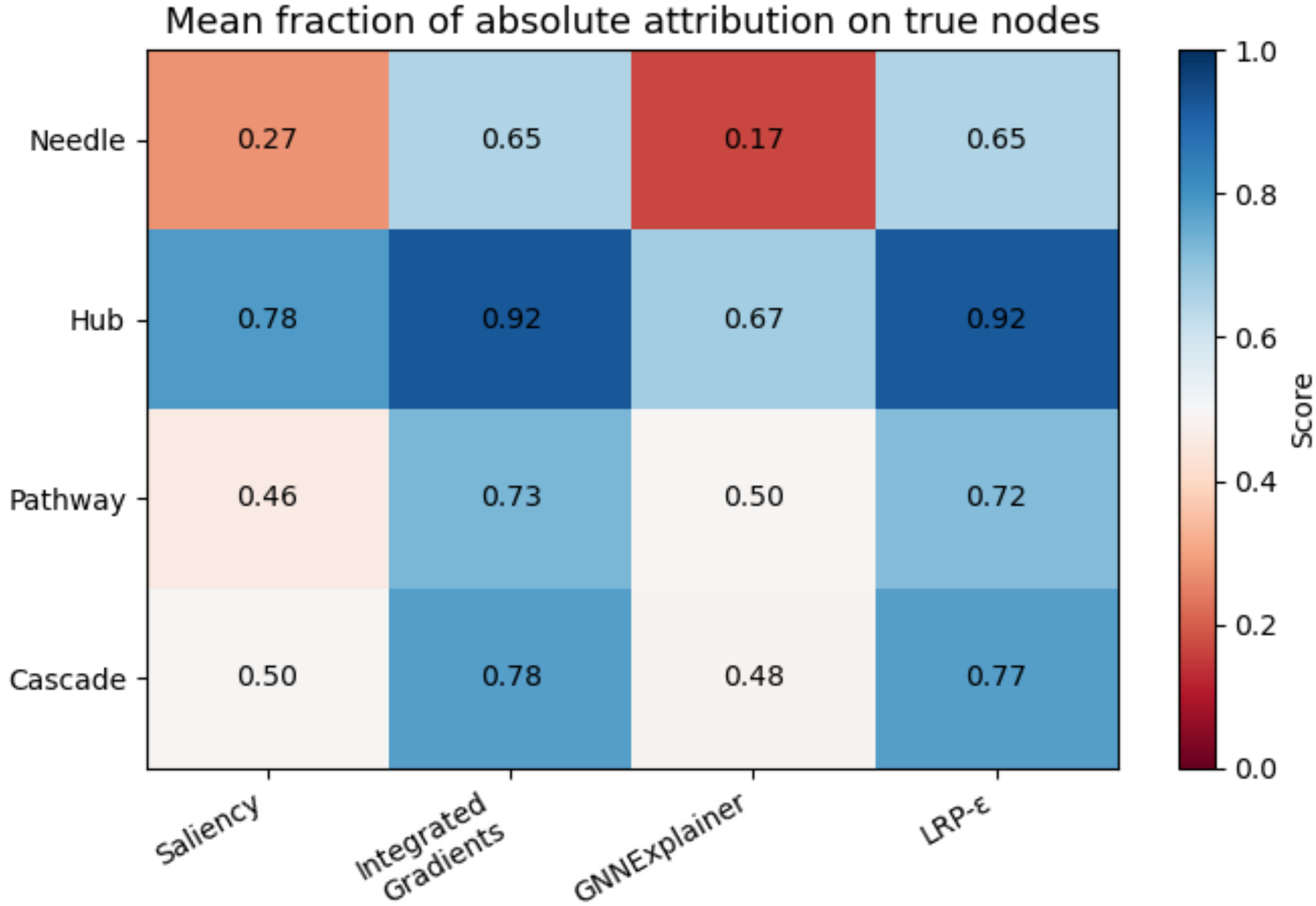

**Figure 2. Quantitative comparison of attribution concentration on ground-truth signal nodes.** Heatmap showing the mean fraction of absolute attribution assigned to true signal nodes across four synthetic biological signal archetypes (rows: needle, hub, pathway, cascade) and explanation methods (columns). Higher values indicate better alignment between model explanations and the underlying ground-truth signal. IG and LRP-ε consistently achieve the highest attribution concentration, particularly for structured and distributed signals (hub, pathway, cascade), whereas saliency and GNNExplainer show reduced alignment, especially in more complex settings. These results highlight systematic differences in how explainers capture biologically meaningful signal structure.

*Synthetic Pathways/Modules*

In the pathway or module setting, signal is distributed more evenly across a module of interacting nodes. (**Figure 1C**) Here, LRP and IG once again showed the strongest performance, with mean absolute fraction on true nodes of 0.72 for both. SA and GNNExplainer showed notably worse performance, with 0.46 and 0.50 mean absolute fraction on true nodes, respectively. Notably, GNNExplainer maintained perfect top-1 and top-5 hit rates but distributed attribution more diffusely, suggesting it is better at identifying a prediction-preserving subset than the full pathway.

*Synthetic Cascades*

The cascade setting represents sequential and hierarchical biological processes where expression signal also upticks further down the cascade. Notable differences between methods existed in this setting, with LRP and IG achieving high alignment again, and mean absolute fraction on true nodes of 0.77 for LRP and 0.78 for IG. In contrast, SA and GNNExplainer similarly aligned, with much lower values of 0.50 (SA) and 0.49 (GNNExplainer). (**Figure 1D**) Taken together, these results indicate that while all methods can identify some part of the cascade, LRP and IG do a much better job of recovering the fully distributed structure.

*Comparing LRP Rules*

We also investigated the utility of using different LRP rules. Across all synthetic datasets, predictive performance was near-perfect for structured motifs (hub, pathway, cascade; test and balanced accuracy = 1.0), with reduced performance for the needle setting (~0.80), confirming its increased difficulty. Despite identical model accuracy, substantial differences emerged between LRP rules in explanation quality. In structured settings, LRP- ε and LRP- γ consistently outperformed αβ and $z^+$, achieving perfect top-1 and top-k hit rates (1.0) and concentrating most relevance on true nodes (~0.93–0.98), indicating highly faithful recovery of the underlying signal. In contrast, αβ and $z^+$ showed weaker localization, with lower top-1 hit rates (~0.60–0.64) and more diffuse relevance (~0.55–0.67), particularly in cascade and pathway scenarios where signal is distributed across multiple nodes. Even in hub settings, where signal is localized, ε and γ maintained substantially sharper attribution than αβ and $z^+$. In the needle dataset, however, all methods struggled: top-1 hit rates remained low (~0.27–0.33), top-k recovery was modest (~0.69–0.80), and relevance assigned to true nodes was limited (~0.17–0.23), indicating that none of the LRP variants reliably capture highly sparse signals. Overall, these results show that LRP-ε and LRP-γ provide the most faithful explanations for structured motifs, while all methods remain challenged by sparse, needle-like patterns. (For full results, see **Supplementary Table ST2**.) For the remainder of the manuscript, where LRP is mentioned it should be taken as equivalent to LRP-ε.

*Evaluation of explainers on real breast cancer motifs*

To evaluate the relative ability of our explainers to attribute relevance to known disease motifs in real data, we employed The Cancer Genome Atlas (TCGA)'s BRCA 1222 sample breast cancer RNA-seq dataset. Combining with the STRING PPI, we built a cancer classifier using a graph convolutional architecture (the same as was used in synthetic examples), which had a balanced validation accuracy of 89.7%. This model was frozen and used for the proceeding explainer analysis.

*Integrated Gradients localizes strong attribution to true breast cancer hubs*

Across all seed genes, IG consistently showed the strongest localization of attribution within hub neighborhoods. Enrichment ratios ranged from 4.4 in BRCA1 and 4.8 in TP53 to 10.7 in ESR1 and 11.2 in MYC. (**Figure 3**) This was accompanied by an extremely high overlap of top-50 most important genes with disease hubs, including 10 for BRCA1, 11 for ESR1, 19 for TP53, and 24 for MYC. (Note that some genes in the 1-hop neighborhood are shared between these hubs). AUC was ~0.3 for all

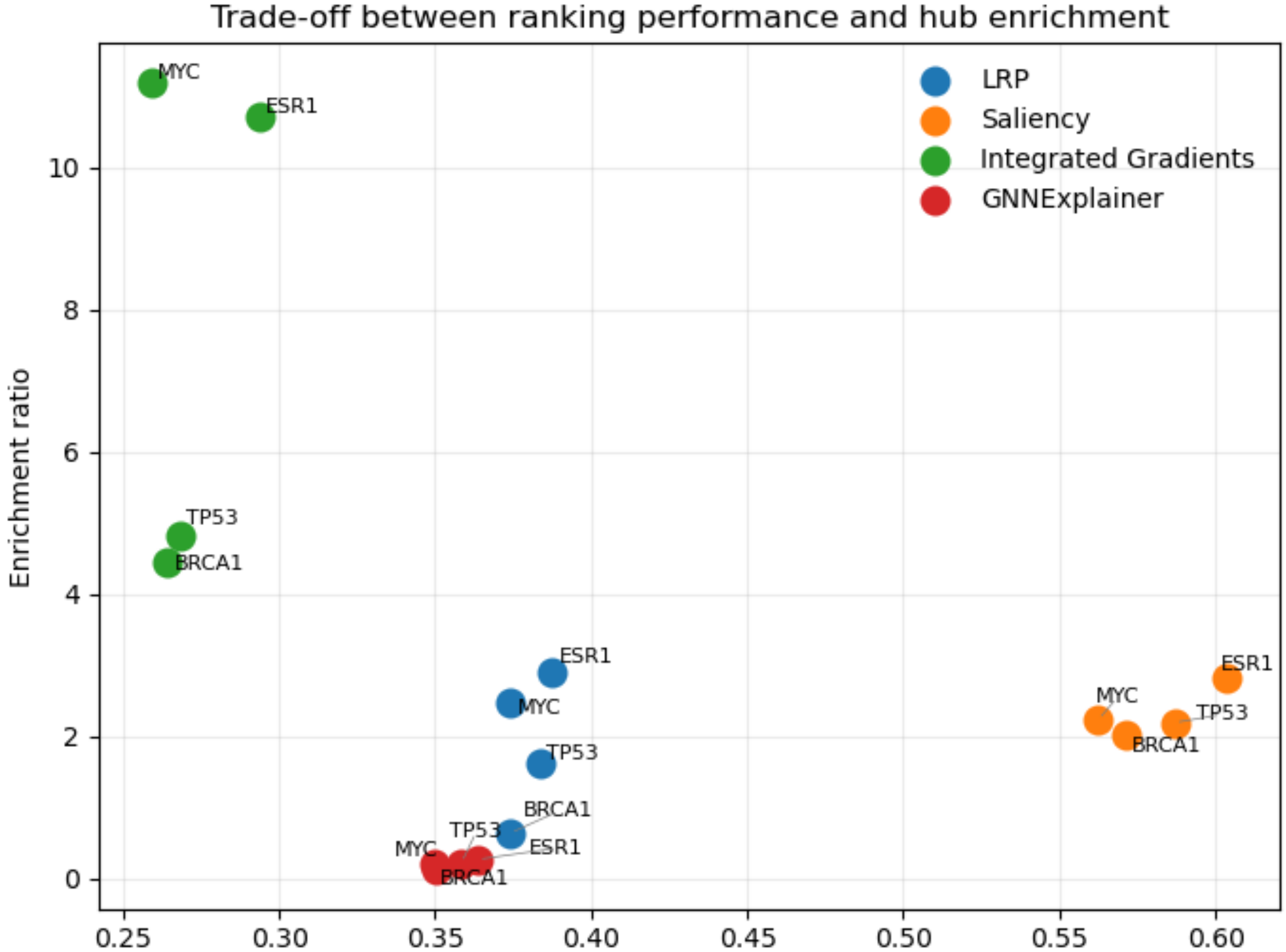

**Figure 3 Trade-off between global ranking performance and hub localization in breast cancer hubs.**

Each point represents the performance of an explanation method for a given seed gene (TP53, BRCA1, ESR1, MYC), with the x-axis showing the ability to globally rank hub-associated nodes (AUC) and the y-axis showing enrichment of attribution within the local hub neighborhood (enrichment ratio). IG achieves the highest enrichment ratios, indicating strong localization of attribution within the hub and its immediate neighborhood, but exhibits lower AUC values, reflecting weaker global ranking performance. In contrast, SA attains the highest AUC, indicating better global discrimination between hub and non-hub nodes, but with reduced enrichment. LRP demonstrates intermediate behavior, balancing localization and ranking performance, while GNNExplainer shows low enrichment and moderate AUC, consistent with more diffuse attribution patterns. These results highlight a trade-off between capturing localized topological influence within hub structures and achieving strong global ranking performance across the graph.

genes. These results indicate that IG concentrates attribution within biologically relevant network regions, effectively identifying the local interaction neighborhood with the seed gene.

*Saliency produces diffuse hub attribution with higher global ranking performance*

In contrast, SA exhibited more moderate enrichment (approximately 2-fold across all genes) but consistently showed the highest AUC among explainer methods, with ~0.6 for all hub neighborhoods. This suggests that while SA assigns higher scores to hub nodes relative to background, it distributes attribution more broadly in the graph. Consequently, SA provides a smoother global ranking of node importance but at the cost of reduced localization in the hub neighborhood.

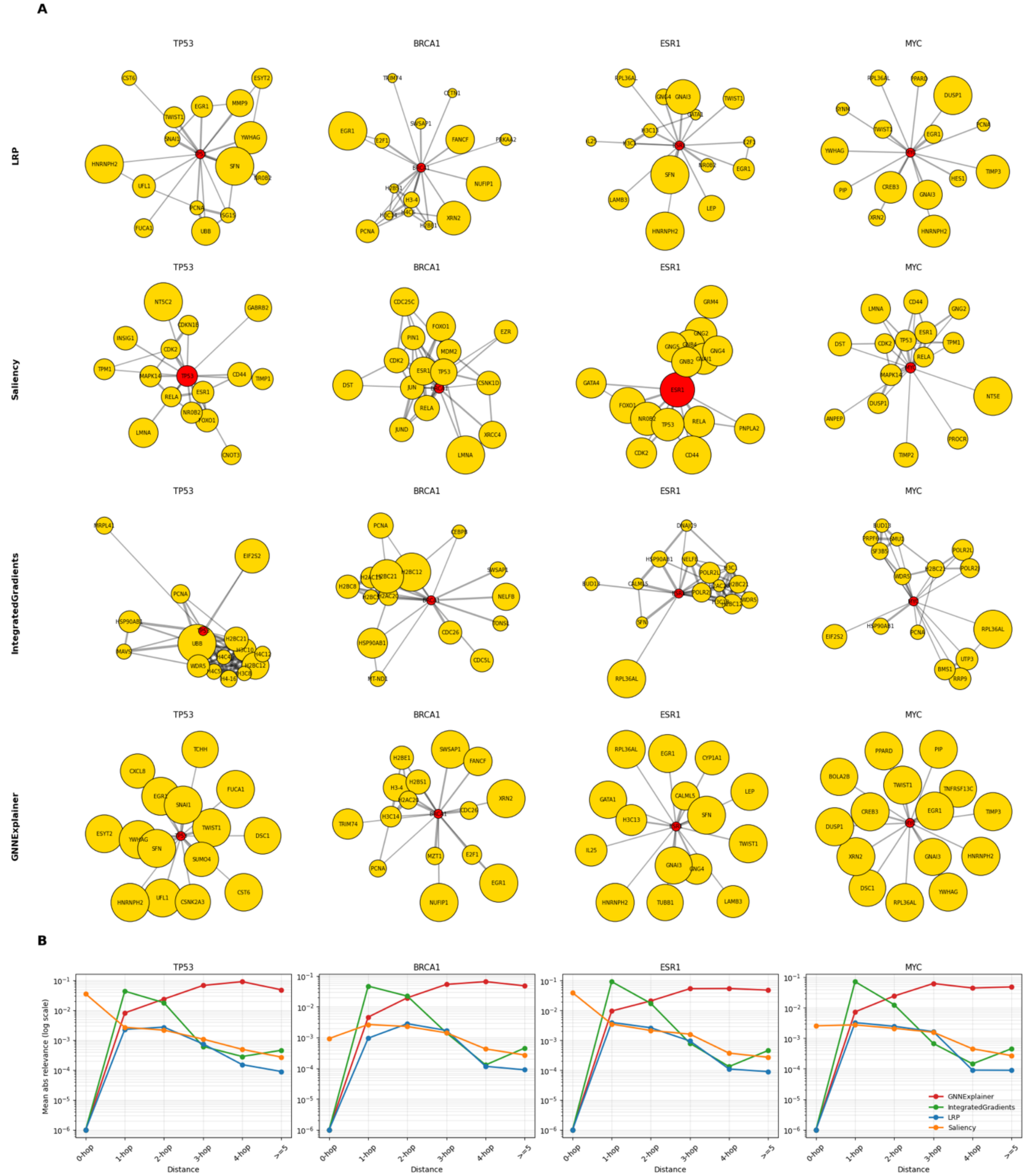

**Figure 4 Explainer Analysis of Breast Cancer Hubs.**
**A** Top 15 one-hop neighbors of each breast cancer hub seed gene (TP53, BRCA1, ESR1, and MYC) ranked by mean absolute attribution for each explanation method. The seed gene is shown in red and neighboring nodes in gold; node size is proportional to mean absolute attribution within each panel. Edges represent STRING protein–protein interactions. **B** Mean absolute attribution as a function of shortest-path distance from each seed gene in the STRING network. Values are shown on a log scale. Integrated Gradients concentrates attribution strongly within the immediate neighborhood of the seed gene, whereas Saliency and LRP show broader spread, and GNNExplainer distributes attribution more diffusely across graph distances.

*LRP shows intermediate behavior on hubs*

LRP demonstrated intermediate performance, with enrichment ratios between 1.6 (TP53) and 2.9 (ESR1) and a moderate top-50 overlap, with highest of 9 for TP53 and MYC, and lowest of 1 for BRCA1. AUC was also high, with values ranging from ~0.4, higher than IG though lower than SA. While LRP was capable of capturing hub-associated signal, it was less effective than IG at concentrating attribution within the hub neighborhood.

*GNNExplainer fails to recover hub structure*

GNNExplainer showed consistently low enrichment ratios (<0.3 in all cases), indicating that attribution was not preferentially concentrated within hub neighborhoods. AUC values remained a modest ~0.4 and a modest 3-9 gene top-50 overlap was observed. Overall, this suggests that other explainers may be better for this application. (For full results, see **Supplementary Table ST3.**)

*Attribution peaks in the neighborhood rather than center of breast cancer hubs*

Across all four hub genes, attribution exhibited a consistent radial structure (except in SA) in which importance was not maximized in the central hub node but instead peaked in its immediate neighborhood (1-hop) before decaying with increasing graph distance (2-4 hops). (**Figure 4B**)For IG, attribution was strongly concentrated at 1-hop nodes (e.g., BRCA1: 0-hop = 0, 1-hop = 0.047, 2-hop = 0.022, 3-hop = 0.0014, 4-hop = 0.00013; ESR1: 0, 0.093, 0.017, 0.0008, 0.00013; MYC: 0, 0.072, 0.013, 0.00067, 0.00014; TP53: 0, 0.043, 0.018, 0.00061, 0.00028), followed by a sharp decay across increasing distances, with values approaching zero beyond 3-4 hops. LRP showed a similar but less sharply peaked profile, with modest attribution at 1-2 hops (e.g., BRCA1: 0, 0.0010, 0.0028) and gradual decay thereafter. In contrast, SA sometimes assigned high attribution directly to the seed node (e.g., ESR1: 0-hop = 0.040, TP53: 0.035), but exhibited weaker localization overall, with relatively shallow decreases across 1-4 hops. This again demonstrates a tendency shown by synthetic examples to identify the central node as highest importance (which may be desirable in certain applications). GNNExplainer displayed a markedly different pattern, with attribution increasing with distance from the seed (e.g., BRCA1: 0, 0.0046, 0.0196, 0.0536, 0.0657), peaking at 3-4 hops, indicating substantial diffusion of importance away from the hub.

Together, these results reveal a consistent “eye-of-the-storm”-like profile for IG and LRP, in which attribution is reduced at the central hub (0-hop), elevated in the immediate neighborhood (1-hop), and subsequently decays with distance (2–4 hops). On the other hand, SA often identifies the central node as most important while showing a less steep decline across more distant shells. Concurrently, the above comparison of enrichment ratio and AUC revealed a clear trade-off between localization of attribution and global ranking performance, demonstrating that for hubs different strengths lie in different methods. This lead us to the hypothesis that a combination of explainer methods may be better for recovering hub structures from real data. To investigate this, we developed a method for agnostic hub identification using both individual explainers and a consensus of them (including SA, IG, and LRP) and observed how well they identified these four disease hubs.

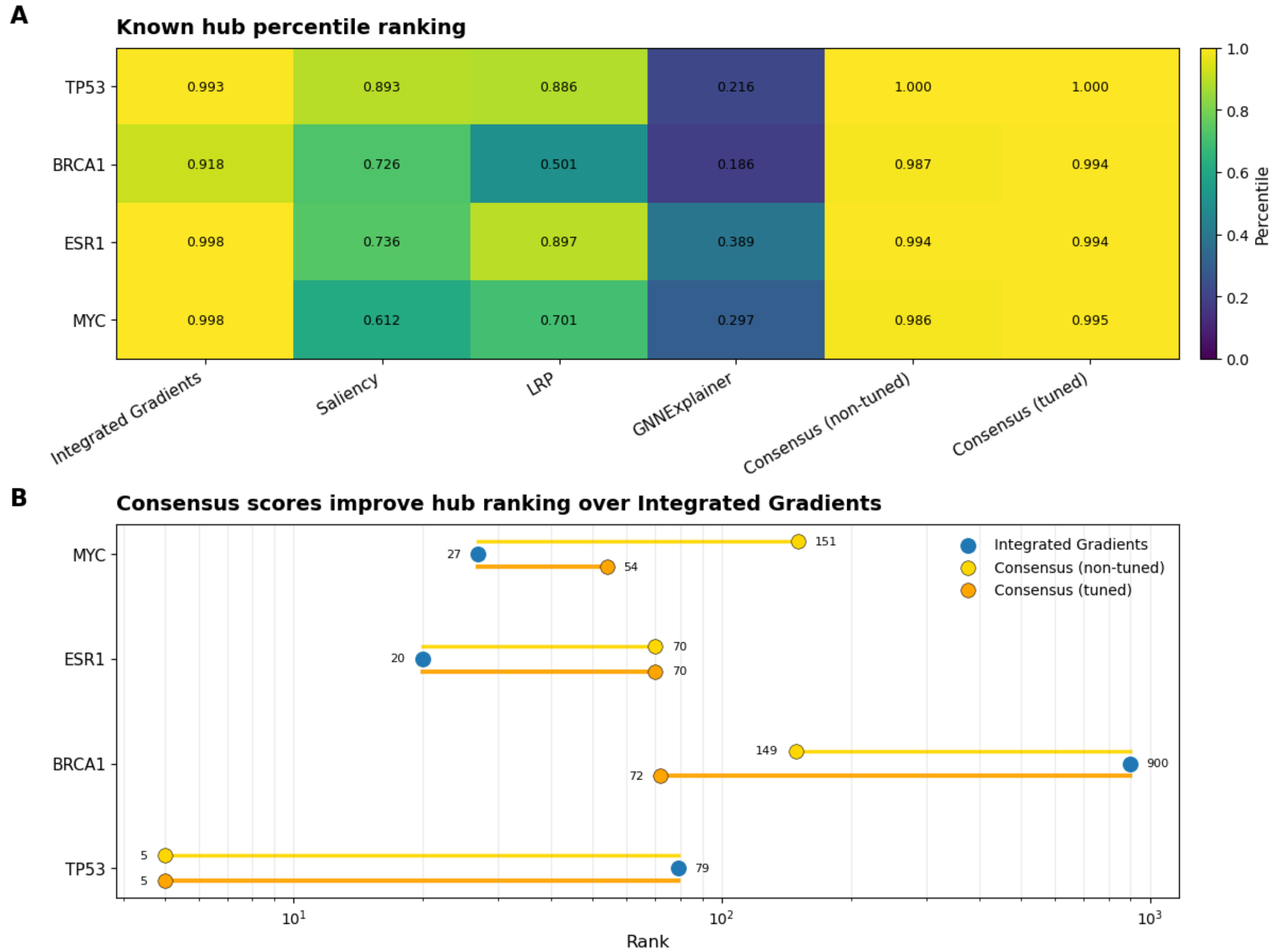

**Figure 5 Consensus attribution improves identification of canonical breast cancer network hubs.** **A** Percentile ranking of four canonical breast cancer-associated hub genes (*TP53*, *BRCA1*, *ESR1* and *MYC*) across explanation methods, evaluated over the full STRING interactome. IG consistently ranks these hubs highly, whereas SA and LRP show variable performance and GNNExplainer performs poorly. In contrast, both non-tuned and tuned consensus scores achieve near-maximal percentiles across all four genes, indicating robust and consistent prioritization. **B** Comparison of global hub ranks (log scale; lower is better) relative to IG. Lines connect ranks assigned by IG (blue) to those from the non-tuned (gold) and tuned (orange) consensus scores. Consensus scoring consistently improves or declines, but by a relatively smaller amount, ranking performance across all four genes, with the tuned consensus yielding the most stable gains. These results demonstrate that aggregating attribution signals enhances recovery of disease-relevant network hubs beyond individual methods.

*Consensus of explainer methods greatly improves agnostic hub identification*

To evaluate the utility of explainer methods in identifying important disease hubs agnostically, we developed a shell-based hub score, which emphasizes the 1-2 hop peak, then ≥3 hop decay in importance score seen above (See *Methods*). We perform a scan of all STRING proteins and apply the shell score for each individual and consensus method. Overall, IG showed highest performance

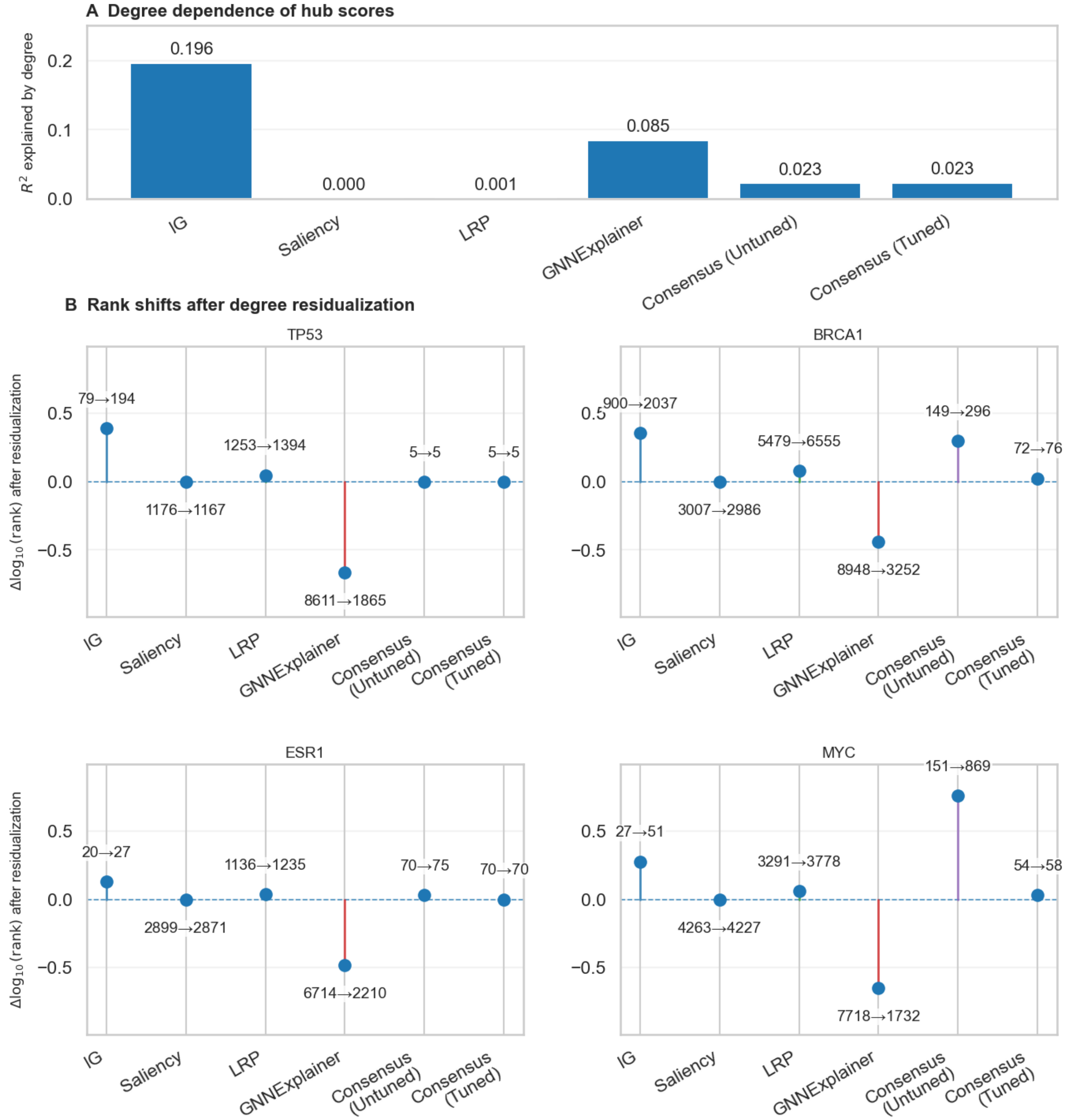

**Figure 6 Degree confounding of hub-centered explanation scores and its impact on prioritization of canonical cancer genes.**

**A** Fraction of variance in hub scores explained by node degree ($R^2$) for each explanation method. IG exhibits strong degree dependence ($R^2 = 0.196$), whereas GNNExplainer shows moderate dependence ($R^2 = 0.085$). In contrast, SA and LRP show negligible dependence, and the consensus hub scores (untuned and tuned) substantially reduce degree confounding ($R^2 = 0.023$). **B** Change in rank of known cancer hub genes (TP53, BRCA1, ESR1, MYC) after residualizing hub scores with respect to node degree. Points represent the change in log-rank ($\Delta\log_{10}(\mathrm{rank})$), with labels indicating raw rank to degree-residualized rank. Positive values indicate improved ranking after removing degree effects, whereas negative values indicate loss of rank. GNNExplainer exhibits large negative shifts across all genes, indicating strong reliance on degree, while Integrated Gradients and LRP show modest changes. Consensus hub scores remain stable across all genes, preserving prioritization of canonical cancer drivers after degree correction.

99.3$^{rd}$, respectively (and ranks at 20, 27, and 79 respectively). (**Figure 5A**) BRCA1 remained lower, in the 91.8$^{th}$ percentile (rank 900). SA and LRP had worse performing scores, but ranked disease hubs with higher scores than average, and much higher than GNNExplainer for TP53 (SA: 89.3$^{rd}$, LRP: 88.6$^{th}$, GNNExplainer: 21.6$^{th}$), BRCA1 (SA: 72.6$^{th}$, LRP: 50.1$^{st}$, GNNExplainer: 18.6$^{th}$), ESR1 (SA: 73.6$^{th}$, LRP: 70.1$^{st}$, GNNExplainer: 38.9$^{th}$), and MYC (SA: 61.2$^{nd}$, LRP: 70.1$^{st}$, GNNExplainer: 29.7$^{th}$).

Next, we produced a rank consensus score for ranking candidate hub genes which is the weighted average of IG rank, SA rank, and LRP rank, with weights of 1, 1, and 0.5 respectively. The ranked consensus outperformed all individual methods, except IG for two genes, MYC and ESR1. For TP53 and BRCA1 the improvement was great, with percentiles increasing from 91.8$^{th}$ to 98.7$^{th}$ (rank 149) for the former and 99.3$^{rd}$ to 100.0$^{th}$ (rank 5) for the latter. (**Figure 5B**) This upside arguably outweighed the downside in ESR1 and MYC, with the former decreasing from 99.8$^{th}$ to 99.4$^{th}$ percentile (rank 70), and latter decreasing from 99.8$^{th}$ to 98.6$^{th}$ (rank 151). These results demonstrate the utility of a consensus rank score in identifying known disease associated hubs.

We were able to further improve the performance of this consensus score by tuning the weights applied to each rank using a grid search, finding optimal weights of 0.69 for IG and 0.31 for SA, and weights 0 for LRP (indicating its signal is possibly redundant to IG) and GNNExplainer (consistent with its poor performance for this task overall). This time improvements were bigger and tradeoffs smaller compared with IG, with TP53 once again improving from 99.3$^{rd}$ to 100$^{th}$ percentile (rank 5), BRCA1 increasing from 91.8$^{th}$ to 99.4$^{th}$ percentile (rank 72), ESR1 decreasing from 99.8$^{th}$ to 99.4$^{th}$ percentile (rank 70), MYC decreasing from 99.8$^{th}$ to 99.5$^{th}$ percentile. These results demonstrate that a rank consensus of explainer methods can be improved via tuning. (For full results, see **Supplementary Table ST4**)

*Degree dependence on explanation scores*

High degree nodes are known to be major drivers of disease, and we therefore expect many genes important in disease to be high degree. To ensure the results of our explainers were being driven by biological signal rather than high degree alone, we performed a degree dependence analysis. Across all methods, hub scores exhibited varying degrees of association with node degree. Rank-based analysis revealed strong relationships for IG (Spearman $r$ = 0.72) and the combined hub scores (Spearman $r$ = 0.69 and 0.79 for untuned and tuned, respectively), indicating that high-degree nodes are preferentially ranked as important. However, to investigate the amount of variance explained by degree, we performed linear modeling. The results demonstrated that degree explained only a limited fraction of variance in hub scores. IG showed the strongest dependence on degree, with $R^2$ = 0.196, indicating that approximately 19.6% of the variance in its hub score is explained by log-degree alone. (**Figure 6A**) GNNExplainer showed a weaker by still noticeable dependence ($R^2$ = 0.085). In contrast, LRP and SA had essentially no linear association with degree ($R^2 < 1 \cdot 10^{-3}$ for both). Importantly, both consensus hub-based scores (tuned and untuned) were only weakly associated with degree (both $R^2$ = 0.023), indicating that the consensus strategy significantly reduced the topology-driven bias relative to IG.

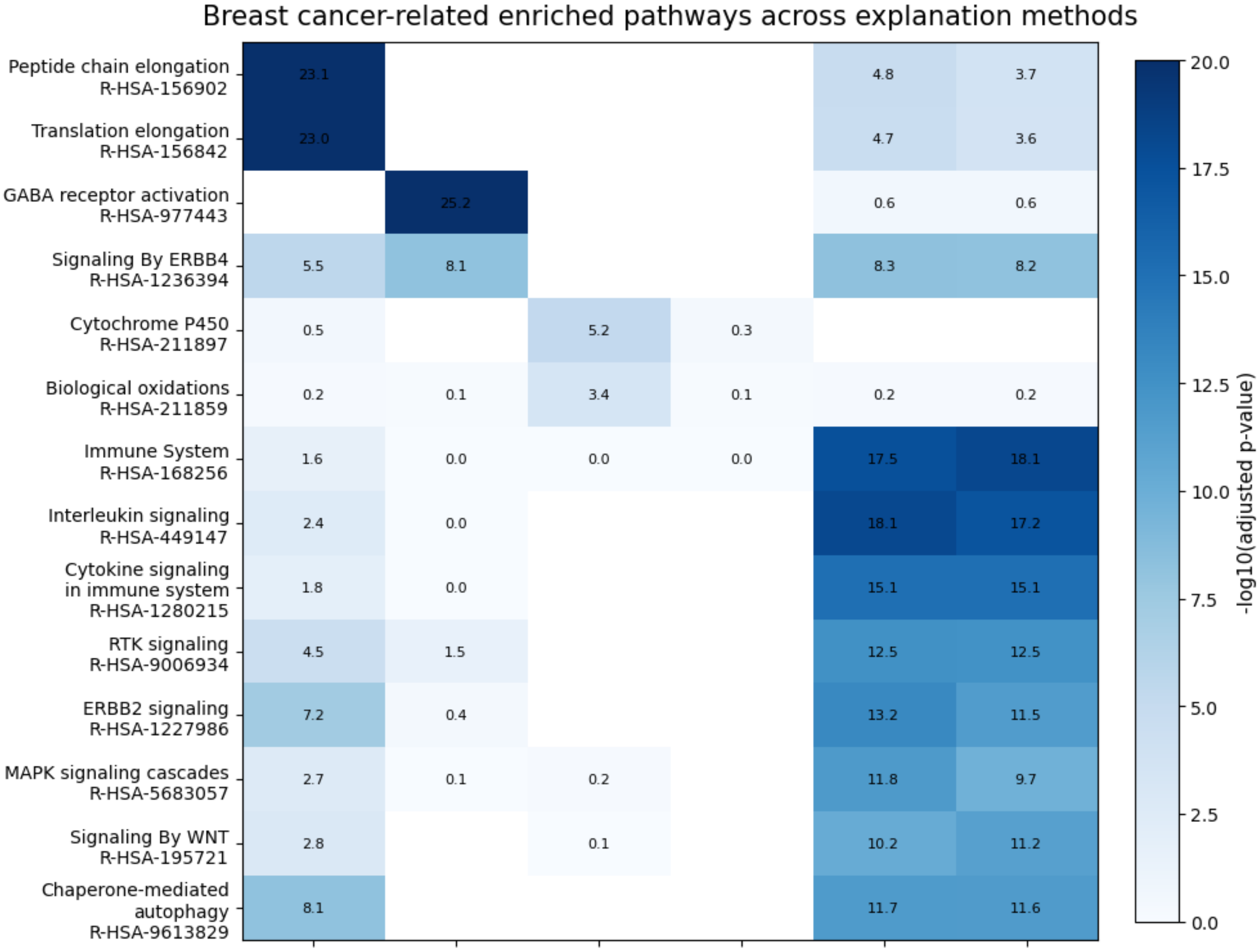

**Figure 7 Pathway enrichment profiles differ substantially across explanation methods, with consensus approaches recovering biologically coherent cancer signaling.**
Heatmap showing enrichment of curated Reactome pathways across top-ranked genes identified by each explanation method. Values represent −log10(adjusted *P*-value) from enrichment analysis of the top 100 candidate hub genes per method. IG is dominated by translational and ribosomal processes (e.g., peptide chain elongation, translation initiation), while SA preferentially highlights neuronal and receptor signaling pathways (e.g., GABA receptor activation). LRP shows modest enrichment of metabolic and detoxification pathways (e.g., cytochrome P450, biological oxidations), while GNNExplainer exhibits minimal or inconsistent enrichment across the curated pathways. In contrast, the consensus hub-based approaches (Consensus Untuned and Consensus Tuned) consistently recover pathways central to cancer biology, including immune signaling (interleukins, cytokine signaling), receptor tyrosine kinase pathways (ERBB2, MAPK), and stress response processes. These results indicate that individual explanation methods capture distinct, and in some cases biologically implausible, signals, whereas consensus approaches better align with known disease-relevant mechanisms.

*Degree confounding analysis for known breast cancer hubs*

Despite this correction, the canonical cancer genes remained highly prioritized after residualization, supporting the conclusion that their signal was not solely attributable to degree. TP53 was particularly stable, remaining ranked 5$^{th}$ by both the tuned and untuned consensus before and after residualization, thus remaining in the 100.0$^{th}$ percentile in both cases. (**Figure 6B**) ESR1 also remained highly ranked, with the tuned consensus preserving a rank of 70 and 99.4$^{th}$ percentile, while untuned consensus changed only minimally from rank 70 to 75. MYC showed somewhat greater attenuation under untuned consensus, dropping from rank 151 to 869, but remained strongly prioritized by the tuned consensus, changing only from rank 54 to 58 and remaining above the 99th percentile. BRCA1 likewise remained highly ranked under the tuned consensus score, shifting only from rank 72 to 76. IG showed moderate degradation after residualization for some genes, for example BRCA1 falling from rank 900 to 2037, whereas GNNExplainer's residualized ranks improved markedly but remained less biologically selective overall, consistent with a noisier and less targeted signal. Together, these results indicate that while some raw explainer scores are partially influenced by node degree, the consensus hub-based scoring framework retains biologically meaningful prioritization even after explicit removal of degree effects.

Together, these results indicate that while node degree strongly influences the ranking of important nodes, biasing explainers towards highly connected regions, degree alone does not explain the magnitude of importance scores, particularly for consensus approaches. This suggests the explanation methods capture signals beyond simple network topology, but operate within a regime where degree imposes a strong baseline structure on node prioritization.

*Pathway enrichment distinguishes disease-relevant explanations from candidate hubs*

To investigate the biological relevance of other candidate hubs (besides the four hallmark breast cancer ones investigated above), we carried out a pathway enrichment analysis of the top 100 candidate genes. IG was dominated by translation-associated pathways, including peptide chain elongation, translation elongation, translation termination, and related ribosomal processes, indicating a strong preference for broad biosynthetic and proliferative programs (which may really by patterned in such a way to indicate disease) rather than disease-specific signaling. (**Figure 7**) SA highlighted neuronal and synaptic pathways, most prominently GABA receptor activation, neurotransmitter receptor signally, and transmission across chemical synapses, though it was able to recover potentially relevant oncogenic signaling such as ERBB4. LRP produced a more fragmented enrichment profile centered on xenobiotic metabolism, cytochrome p450 activity, biological oxidations, and lipid mediator biosynthesis, suggesting partial capture of metabolic and oxidative stress programs but less coherent disease specificity.

In contrast, the consensus hub scores recovered a substantially more plausible cancer-relevant signaling landscape. The untuned consensus enriched interleukin signaling, cytokine signaling, immune system pathways, cellular stress responses, diseases of signal transduction by growth factor receptors, receptor tyrosine kinase signaling, ERBB2 signaling, MAPK family signaling cascades, and chaperone-mediated autophagy. The tuned consensus strengthened this pattern further, retaining immune and stress-response pathways while additionally highlighting ERBB2, non-receptor tyrosine kinase signaling, WNT signaling, and ERBB2:ERBB3-related pathways. Together, these results indicate that individual explainers sometimes either emphasize generic cellular programs or biologically implausible pathway sets, whereas consensus hub scoring shifts enrichment toward coherent oncogenic and tumor-associated signaling modules that are noticeably more consistent with breast cancer biology. (For full results, see **Supplementary Table ST7**)

## Discussion

Our results show that explanation methods capture distinct and complementary aspects of graph-based biological signal. In controlled synthetic benchmarks, IG and LRP consistently outperform other methods in recovering distributed and structured motifs, such as pathways and cascades, indicating their suitability for identifying multi-node, spatially extended signals. In contrast, SA shows a strong tendency to concentrate attribution on central nodes, particularly in hub-like settings, suggesting sensitivity to dominant or high-magnitude features.

We hypothesize that the observed differences among explanation methods arise from the distinct explanatory objectives they optimize. SA emphasizes local sensitivity, which is advantageous for sparse single-node signals but less effective for distributed biological programs. GNNExplainer seeks a compact, sufficient explanation through mask optimization, which may recover a prediction-preserving subgraph without capturing the full set of biologically relevant contributors. In contrast, IG and LRP better capture distributed evidence across hubs, modules, and cascades, making them more suitable for faithful, pathway-oriented biological interpretation. This distinction provides a mechanistic explanation for the observed differences in both synthetic and biological settings.

In real-world breast cancer data, we identify a consistent and previously underappreciated topological pattern in explanation outputs. For known disease-associated hubs, attribution is not maximized at the central node itself but instead peaks in the immediate 1-hop neighborhood and decays progressively across more distant nodes. This "eye-of-the-storm" structure is most pronounced for IG and LRP, which also show the highest enrichment within known hubs.

This finding has two important implications. First, it suggests that biologically relevant signal in graph models is often distributed across local neighborhoods rather than concentrated at individual nodes. Second, it provides a concrete, topology-aware signature that can be exploited to identify candidate disease hubs independent of prior knowledge.

Motivated by the observed shell-based topology and complementary strengths of explanation methods, we developed two strategies for candidate hub identification. First, we introduce a shell-based hub score that explicitly captures the enrichment of attribution within local neighborhoods relative to background. Second, we construct consensus ranking schemes that integrate attribution signals across multiple explainers.

Consensus approaches consistently improve identification of canonical cancer genes and remain robust after correcting for degree confounding. Notably, optimized consensus scores, placing higher weight on IG and moderate weight on SA, yield the strongest performance, suggesting that combining local enrichment sensitivity with global ranking ability provides a more complete representation of disease-relevant signal.

As key concern in network-based analyses is confounding by node degree, we investigated this effect in individual and consensus explainers. We find that IG exhibits measurable degree dependence, while consensus scores substantially reduce this effect. Importantly, consensus-based prioritization of canonical cancer genes remains stable after degree residualization, indicating that these methods capture signal beyond simple topological centrality.

Further, we observe at the pathway level that consensus approaches recover biologically coherent cancer programs, including ERBB2, RTK, and MAPK pathway signaling along with immune and cytokine signaling which indicates tumor microenvironment and inflammation responses. In contrast, individual explainers such as IG frequently highlight pathways that are less directly related to breast cancer biology, such as general translation or housekeeping processes. Though these processes may also have fine structure relevant to disease, we consider identification of bona fide disease pathways preferrable in our explanation. These results further support the biological relevance of integrating explanation signals.

## Conclusion

Together, our findings suggest that graph explanations in biological systems should be interpreted not as single-method outputs, but as structured signals shaped by both network topology and method-specific biases. By combining synthetic ground-truth benchmarks with TCGA breast cancer analysis, we show that different explainers recover different graph signal geometries and that topology-aware consensus scoring can improve the prioritization of canonical cancer-associated genes and processes. These results provide a practical framework for hypothesis generation from graph-based omics models, while highlighting the need for model-sensitivity analyses, degree-matched controls, and independent biological validation before mechanistic conclusions are drawn.

## Methods

### *Synthetic biological signal patterns/motifs*

To assess graph explanations under controlled and biologically interpretable conditions, we constructed four basic motif types designed to mimic four distinct molecular signal organizations found in PPIs. For each motif type, we trained a benchmark encoding a predefined signal pattern within a fixed graph topology and assigned binary labels to represent whether the signal (represented in node features) was up- or down-regulated. This design provided explicit ground truth signal nodes, enabling direct evaluation of signal recovery by each explanation method.

For each benchmark, each graph was represented as a Pytorch Geometric Data object, with node features and undirected edges. Node features were sampled from zero mean gaussian noise with lower-level variance, $x \sim \mathcal{N}(0, 0.3^2)$, and signal was introduced in a single feature dimension for the predefined node subset. For these ground truth experiments, class labels were balanced and signal nodes were perturbed directly according to class label, ensuring the predictive rule was transparent and that explanation performance primarily reflected the ability to recover the known motif.

In the formulation below, let $x_{i,f}$ denote feature $f$ at node $i$. Binary labels were generated from a benchmark-specific score $s$, with positive graphs corresponding to $s > 0$ in the general formulation.

*Needle*

The needle benchmark represents a sparse “needle-in-the-haystack” setting in which the graph label depends on a single decisive node-feature pair. Graphs contained 8 nodes total and 1 feature per node. The label was determined by the value of the feature in a single target node, with score

$$s_{needle} = x_{i,f}.$$

Here positive-class graphs received a strong positive perturbation and negative-class graphs received a strong negative perturbation. This benchmark tested whether a single dominant causal node could be isolated by each explainer method.

*Hub*

The hub benchmark modeled a signaling hub and its immediate neighborhood. Graphs contained 10 nodes and 1 feature per node. The signal was comprised of a central hub node and its 1-hop neighbors, with the hub itself perturbed moderately and the surrounding neighbors perturbed more strongly. The score to determine label in this case was defined as:

$$s_{hub} = 0.5x_{i,f} + 1.2 \sum_{n \in \mathcal{N}_i} x_{n,f}$$

Where $\mathcal{N}_i$ is the 1-hop neighborhood of $i$. In this setting, the central hub node was perturbed to $\pm 0.8$ and its neighboring nodes to $\pm 2.0$, plus gaussian noise. It was designed to reflect situations where a signaling protein, such as a TF, shows increased yet lower abundance compared to its signal recipients. This method was intended to determine the ability to recover an entire distributed neighborhood rather than over-concentrating attention on the hub center alone.

*Pathway/Module*

The pathway or module benchmark was meant to represent the coordinated activation of a connected subnetwork. Graphs contained 12 nodes and 1 feature per node, with a predefined connected module of 5 nodes embedded within a large background topology. Labels were determined by the average activity of features:

$$s_{module} = \frac{1}{|M|} \sum_{n \in M} x_{n,f}$$

where $M$ is the 5-node module set. This benchmark tested whether explainers could identify a coherent, spatially contiguous subnetwork corresponding to a pathway-like biological process.

*Cascade*

The cascade benchmark is meant to represent an ordered signaling chain. Graphs contained a total of 12 nodes and 1 feature per node, with five predefined cascade nodes embedded in a background graph. In positive examples, signal increases progressively along the cascade with weights $w = [0.5, 0.8, 1.0, 1.2, 1.5]$, whereas negative examples showed the opposite pattern. The score to determine label was:

$$s_{cascade} = \sum_{j=1}^{5} w_j x_{C_j,f}$$

Where $C$ is the cascade node set and $w$ are our weights, described above. This benchmark tested whether our explainers could recover structured signal distributed along a directional chain rather than collapsing attribution to a single endpoint or bottleneck.

*Graph Neural Network Architecture and Training*

For all synthetic benchmarks, we trained the same standard graph classification architecture to enable direct comparison of explanation methods across signal types. The model consisted of three graph convolution layers with additive neighborhood aggregation, followed by a ReLU nonlinearity transformation. The final node embeddings were aggregated using global mean pooling, and graph-level logits were produced by a linear classification layer.

Let $G = (V, E)$ be a graph with node features $x_i \in \mathbb{R}^F$, where $F$ is the number of feature dimensions. For a single graph convolution layer, messages were aggregated additively from source nodes to destination nodes:

$$m_i = \sum_{j \in \mathcal{N}(i)} x_j$$

And combined with a root-node contribution:

$$z_i = W_{neighbor} m_i + W_{root} x_i + b$$

The hidden representation after nonlinearity:

$$h_i = \mathrm{ReLU}(z_i)$$

After applying this update for three successive layers, node embeddings at the final layer were pooled by global mean pooling:

$$g = \frac{1}{|V|} \sum_{i \in V} h_i^{(3)}$$

Graph-level logits were then computed as:

$$\ell = W_{fc}g + b_{fc}$$

Where $W_{fc}$ and $b_{fc}$ are the fully connected weights and biases, respectively. The model was trained using cross-entropy loss with Adam, learning rate of $10^{-3}$, weight decay of $10^{-4}$, hidden dimension 32, batch size 16, and 40 training epochs. Data (300 synthetic graphs to each benchmark) was split into training, validation, and testing sets in ratio 75:15:15, with the highest validation accuracy model being retained.

*Explainer implementation*

We compared four explanation methods: SA, IG, GNNExplainer, and LRP (using $\varepsilon$-rule). SA and IG were implemented using the Captum-based explanation interface in Pytorch Geometric with node attribute masks as the explanation target. GNNExplainer was implemented using the Pytorch Geometric Explainer framework with 100 optimization epochs. For these methods, node-level attribution scores were feature-level attributions for the single feature nodes.

LRP was implemented explicitly for the graph architecture used in this study. Relevance was initialized at the predicted class logit and propagated backward through the linear classifier, global mean pooling, ReLU nonlinearities, and graph convolution layers. Within each graph convolution block, relevance was redistributed across the root branch and neighborhood aggregation branch in proportion to their stabilized contributions and then further redistributed from aggregated messages back to source nodes. Linear layers used the $\varepsilon$-stabilized redistribution rule, and global mean pooling was inverted by redistributing graph-level relevance back to node embeddings in proportion to their contributions to the pooled representation.

*Quantitative evaluation of explanation quality*

Because the signal-generating nodes were known by construction for each synthetic graph explanation quality could be evaluated directly against ground truth For each, explanation method and benchmark, we computed four complementary metrics on 20 held-out test graphs, top-1 hit rate, top-5 hit rate, mean fraction of absolute attribution on true nodes, and mean fraction of positive attribution on true nodes. Top-1 hit rate was defined as the proportion of graphs for which the single highest-scoring node belonged to the true signal node set. Top-5 absolute hit rate was defined as the proportion of graphs for each at least one true signal node appeared among the top five, ranked by absolute attribution magnitude. Mean fraction of absolute attribution on true nodes measured the fraction of total absolute node attribution assigned to the ground-truth signal nodes, averaged across graphs. This metric quantified how well a method concentrated explanatory mass on true generating structure, irrespective of sign. Mean fraction of positive attribution on true nodes measured the fraction of total *positive* node attribution assigned to the ground-truth signal nodes, averaged across graphs. This metric emphasized whether a method assigned supportive, class-consistent relevance rather than distributing positive importance elsewhere.

*Comparison of LRP Rules*

To assess the impact of different relevance propagation rules, we compared the performance of variants $\varepsilon$-rule, $z^+$-rule, $\gamma$-rule, and $\alpha\beta$-rule. All rules were tested on a fixed model across comparisons so that the only differences in explanation would originate from the propagation rules. In all cases, relevance was initialized at the class logit and propagated backwards through the network. For a linear layer with input $x$, weights $W$, and output relevance $R^{(l+1)}$, the $\varepsilon$-rule redistributes relevance according to

$$R^{(l)} = x \odot \left( \frac{R^{(l+1)}}{z + \epsilon \cdot sign(z)} W \right),$$

Where $z = W^{\top}$ and $\epsilon$ is a small stabilizing constant. The $z^+$-rule restricts the propagation to positive contributions by using only $W^{+} = \max(W, 0)$ and $x^{+} = \max(x, 0)$, while the $\gamma$-rule amplifies positive contributions via $W + \gamma W^{+}$. The $\alpha\beta$-rule separates positive and negative contributions and redistributes them with coefficients $\alpha$ and $\beta$, such that $\alpha - \beta = 1$.

All rules were evaluated on the same held-out test graphs across all synthetic benchmarks. Explanation quality was quantified using the same metrics as described above, including Top-1 hit rate, Top-5 absolute hit rate, and the fraction of attribution assigned to ground-truth signal nodes. This setup enabled direct comparison of how different LRP formulations affect the localization and distribution of relevance in graph-based models.

*TCGA BRCA RNA-seq disease motif explainer analysis*

To observe the relative ability of explainers to identify disease-relevant localized importance in real data, we investigated motifs in The Cancer Genome Atlas (TCGA)'s BRCA breast cancer dataset. In total we used 1102 log-transformed transcript-per-million (TPM) samples, divided into 1109 cancer and 113 normal adjacent tissue samples. RNA-seq values were projected onto the STRING PPI, with edges reduced to 100,000 highest confidence score from physical experiments for computational efficiency (rather than simulated evidence also included in the project). Classifiers were built using the same three-convolution architecture described above for the synthetic motifs to classify cancer from normal adjacent tissue samples, except with a weighted binary cross entropy loss function to account for class imbalance. The data was split into training, validation, and test sets in a ratio of 80:10:10. Explainers were run on the same trained model to ensure that any differences in explanation were only due to methodological differences in the explainers.

*Hub-based evaluation of explainer methods*

To assess whether explainers recover biologically meaningful hub structures, we evaluated their ability to locally attribute importance in the neighborhood of known cancer driver genes TP53, BRCA1, ESR1, and MYC. For each gene, we defined a hub neighborhood as the full 1-hop subgraph in STRING consisting of seed protein and any directly interacting proteins. Node-level attribution scores were computed for each method and aggregated across samples to obtain mean absolute relevance score per gene. For each gene and method, genes were labeled as belonging to either the hub neighborhood or background.

We quantified the degree to which each method concentrated attribution within the hub using an enrichment ratio, defined as:

$$Enrichment = \frac{mean\ attribution\ inside\ hub}{mean\ attribution\ outside\ hub}$$

An enrichment ratio greater than 1 indicates preferential allocation of importance to the hub neighborhood. To assess whether highly ranked nodes corresponded to the hub, we computed the top-50 overlap, defined as the number of hub nodes appearing among the 50 highest-scoring nodes for each method. Finally, we also computed the area under the receiver operating characteristic curve (AUC), treating hub membership as the positive class and attribution score as the ranking signal. While AUC reflects global ranking performance across all nodes it does not directly capture localization of attribution within biologically relevant signals.

*Hub distance-based attribution analysis*

To quantify how attribution propagates across network topology, we computed the mean absolute attribution as a function of shortest-path distance from each seed gene (TP53, BRCA1, ESR1, MYC) in the STRING protein–protein interaction network. For each method, node-level attributions were first averaged across samples to obtain a gene-level importance score. Shortest-path distances from each seed protein were then computed using NetworkX, and nodes were grouped into discrete distance bins (0-hop, 1-hop, 2-hop, 3-hop, 4-hop, and ≥5-hop). For each bin, we calculated the mean absolute attribution across all nodes within that distance group.

Distances were computed up to four hops from the seed node, with all remaining nodes grouped into a ≥5-hop category. This analysis enables assessment of how attribution is spatially distributed relative to the central hub and its surrounding neighborhood, providing a quantitative measure of localization versus diffusion of signal across the graph.

*Consensus hub scoring*

To identify candidate hub-centered signaling structures across the STRING protein–protein interaction network, we first aggregated gene-level attribution scores from four explanation methods: IG, SA, and LRP. For each method, we loaded a table of mean absolute relevance scores per protein and retained only proteins present in the STRING graph. Missing values were set to zero. Because raw attribution magnitudes differ substantially across explanation methods, each method was percentile-rank normalized across all scored proteins before combination.

Motivated by the observed superior performance of IG for enrichment of disease hubs, the complimentary high rank identification AUC performance of SA, and the modest performance of LRP (by both metrics), we constructed an initial consensus hub-oriented score. We combined the percentile-normalized Integrated Gradients, Saliency, and LRP scores, with reduced weight assigned to LRP:

$$ConsensusHub = \frac{IG_{rank} + SA_{rank} + 0.5LRP_{rank}}{2.5}$$

To evaluate the utility of optimizing the weight of each explainer individually, a tuned consensus score was then optimized by grid search over non-negative weights assigned to Integrated Gradients, Saliency, and LRP:

$$OptimizedConsensus = w_1 IG_{rank} + w_2 SA_{rank} + w_3 LRP_{rank} + w_4 GNNEx_{rank}$$

*Shell-based hub score*

To quantify whether a seed protein sat at the center of a locally enriched signaling neighborhood, we defined concentric graph-distance shells around each node in STRING. For each candidate seed protein, shortest-path distances were computed and nodes were partitioned into four groups: the seed itself (0-hop), immediate neighbors (1-hop), second-order neighbors (2-hop), and a distant background (≥3 hops). For each explanation method, we computed the mean attribution in each shell.

A hub score was then defined as the ratio of local neighborhood attribution to distant background attribution:

$$HubScore = \frac{mean(1 - hop) + mean(2 - hop)}{2 \cdot mean(\geq 3 - hop)}$$

This score was designed to capture the "eye-of-the-storm" profile observed in graph explanations, in which attribution is attenuated at the central node but elevated in its immediate neighborhood. Thus, high hub scores indicate proteins whose local 1–2 hop neighborhoods are preferentially enriched for attribution relative to the broader interactome. For the tuned consensus score, shell-specific percentile-normalized values from IG, SS, and LRP were linearly combined using optimized weights, and the same shell-based hub score was recomputed.

*Degree Confounding Analysis*

To assess whether hub scores were driven by network topology rather than disease-relevant signal, we quantified the relationship between each method's hub score and node degree across all candidate seed proteins in the STRING protein-protein interaction graph. For each explanation method, we computed both Spearman rank correlation and Pearson correlation between node degree (or log-transformed degree) and the corresponding hub score. Spearman correlation was used as the primary measure to assess monotonic relationships between degree and importance ranking. For each explainer-derived hub score, we regressed against log-degree using ordinary least squares:

$$HubScore = \beta_0 + \beta_1 \log(degree) + \varepsilon$$

Pearson correlation with coefficient of determination $R^2$ was used to determine the fraction of variance explained by degree alone. Higher $R^2$ therefore indicates stronger degree confounding.

We next performed degree residualization to determine whether the prioritization of canonical breast cancer genes persisted after removing the linear contribution of degree. For each scoring method, residual hub scores were obtained as the difference between the observed hub score and the fitted value of the regression on log-degree. Genes were then re-ranked by both their raw hub score and their degree-residualized hub score. We specifically evaluated TP53, BRCA1, ESR1, and MYC once again, comparing their raw and residualized ranks and percentiles. This allowed us to test whether the performance of these genes was a consequence of their centrality in the network or reflected signal beyond degree.

*Pathway enrichment analysis of candidate hubs*

To assess the biological programs captured by hub-centered explanation scores, we performed pathway enrichment analysis on the top-ranked candidate hub genes identified by each explanation method. For each method, genes were ranked by hub score and the top 100 candidate hub genes were selected. These gene sets were analyzed using Enrichr via the gseapy interface against the Reactome 2022 pathway database for *Homo sapiens*. Pathways were ranked by adjusted p-value, and enrichment significance was summarized as $-\log_{10}$ adjusted p-value. To enable comparison across methods, we visualized a curated subset of representative pathways spanning the dominant biological themes recovered by the individual explainers and consensus scores, including translation-associated pathways, neuronal and receptor-signaling pathways, metabolic and oxidative pathways, and cancer-relevant immune and growth factor signaling programs.

**Funding:**

This work was supported by several grants, including the AIDA project, funded by UK Research and Innovation (Grant No. 10058099) and the European Union (Grant No. 101095359); an Imperial College President's Scholarship; and the Cancer Research UK Early Detection and Diagnosis Programme scheme (Grant Ref. EDDPGM-May21\100007).

**Code Availability Statement:**

Code is available upon request by contacting corresponding author and will be made fully available to public bitbucket repository upon peer-reviewed publication.

**Author contributions:**

K.H. was responsible for methodological development, simulation design and experiments, evaluation of results, and manuscript preparation. I.L. and D.V. contributed to the conceptualization, methodology development, and manuscript review. K.V. supervised the methodological development, simulations, and evaluations, and was responsible for securing funding and overall project management. All authors contributed to the writing and editing of the manuscript and approved the final version.

| Dataset | Method | Top-1 hit rate | Top-5 abs hit rate | Abs. relevance on true nodes | Pos. relevance on true nodes |
|---|---|---|---|---|---|
| Needle | Saliency | 1 | 1 | 0.274 | 0.274 |
| Needle | Integrated Gradients | 0.7 | 1 | 0.649 | 0.599 |
| Needle | GNNExplainer | 0.9 | 1 | 0.167 | 0.167 |
| Needle | LRP-ε | 0.7 | 1 | 0.651 | 0.596 |
| Hub | Saliency | 1 | 1 | 0.772 | 0.772 |
| Hub | Integrated Gradients | 1 | 1 | 0.93 | 0.977 |
| Hub | GNNExplainer | 1 | 1 | 0.676 | 0.676 |
| Hub | LRP-ε | 1 | 1 | 0.924 | 0.978 |
| Pathway | Saliency | 0.7 | 0.9 | 0.461 | 0.461 |
| Pathway | Integrated Gradients | 1 | 1 | 0.727 | 0.904 |
| Pathway | GNNExplainer | 1 | 1 | 0.495 | 0.495 |
| Pathway | LRP-ε | 0.8 | 1 | 0.718 | 0.72 |
| Cascade | Saliency | 0.9 | 1 | 0.496 | 0.496 |
| Cascade | Integrated Gradients | 1 | 1 | 0.776 | 0.869 |
| Cascade | GNNExplainer | 0.95 | 1 | 0.482 | 0.482 |
| Cascade | LRP-ε | 1 | 1 | 0.774 | 0.851 |

**Supplementary Table ST1Performance of Explainers on Synthetic Motifs.** Quantitative comparison of Saliency Attribution, Integrated Gradients, GNNExplainer, and LRP-ε across four synthetic signal motifs: needle, hub, pathway/module, and cascade. Performance is reported using top-1 hit rate, top-5 absolute hit rate, fraction of absolute relevance assigned to ground-truth nodes, and fraction of positive relevance assigned to ground-truth nodes.

| dataset | rule | test_acc | test_bal_acc | top1_hit_rate | topk_abs_hit_rate | mean_frac_pos_on_true | mean_frac_abs_on_true |
|---|---|---|---|---|---|---|---|
| cascade | alphabeta | 1 | 1 | 0.622222222 | 1 | 0.674743221 | 0.593511999 |
| cascade | epsilon | 1 | 1 | 1 | 1 | 0.930012368 | 0.873423436 |
| cascade | gamma | 1 | 1 | 1 | 1 | 0.930692557 | 0.883455297 |
| cascade | zplus | 1 | 1 | 0.6 | 0.866666667 | 0.555814836 | 0.555814836 |
| hub | alphabeta | 1 | 1 | 0.6 | 1 | 0.592609004 | 0.844640574 |
| hub | epsilon | 1 | 1 | 1 | 1 | 0.982003774 | 0.966566082 |
| hub | gamma | 1 | 1 | 1 | 1 | 0.98354759 | 0.97509108 |
| hub | zplus | 1 | 1 | 0.644444444 | 1 | 0.59216366 | 0.59216366 |
| needle | alphabeta | 0.8 | 0.795 | 0.333333333 | 0.733333333 | 0.201375657 | 0.212894356 |
| needle | epsilon | 0.8 | 0.795 | 0.266666667 | 0.688888889 | 0.216700393 | 0.164408008 |
| needle | gamma | 0.8 | 0.795 | 0.333333333 | 0.8 | 0.227729686 | 0.223405755 |
| needle | zplus | 0.8 | 0.795 | 0.266666667 | 0.777777778 | 0.174844775 | 0.22876596 |
| pathway | alphabeta | 1 | 1 | 0.6 | 1 | 0.573565063 | 0.682736721 |
| pathway | epsilon | 1 | 1 | 1 | 1 | 0.938159489 | 0.890079073 |
| pathway | gamma | 1 | 1 | 1 | 1 | 0.941966887 | 0.913444732 |
| pathway | zplus | 1 | 1 | 0.6 | 0.933333333 | 0.569516685 | 0.569516685 |

**Supplementary Table ST2 Performance of Different LRP Rules on Synthetic Motifs.** Evaluation of ε, γ, αβ, and $z^+$ LRP rules across the synthetic benchmark datasets. The table reports predictive accuracy and explanation quality metrics, enabling comparison of how different relevance

redistribution rules affect localization of known ground-truth signal nodes.

| method | seed_gene | hub_size | inside_mean_abs | outside_mean_abs | enrichment_ratio | top50_overlap | expected_top50_overlap | mannwhitney_p | auc | effect_size | p_value |
|---|---|---|---|---|---|---|---|---|---|---|---|
| GNNExplainer | BRCA1 | 775 | 0.004624927 | 0.035538884 | 0.130137096 | 3 | 2.013196176 | 1 | 0.350470018 | -0.511925352 | 1 |
| IntegratedGradients | BRCA1 | 775 | 0.046987429 | 0.010589985 | 4.436968618 | 10 | 2.013196176 | 1 | 0.264170152 | 0.11936038 | 0.017982018 |
| LRP | BRCA1 | 775 | 0.000955142 | 0.00149821 | 0.63752196 | 1 | 2.013196176 | 1 | 0.373897189 | -0.030012022 | 0.711288711 |
| Saliency | BRCA1 | 775 | 0.002655421 | 0.001306198 | 2.032940054 | 7 | 2.013196176 | 5.36E-11 | 0.571327646 | 0.226564442 | 0.000999001 |
| GNNExplainer | ESR1 | 873 | 0.00958524 | 0.035446637 | 0.27041325 | 5 | 2.26776808 | 1 | 0.363586685 | -0.392593534 | 1 |
| IntegratedGradients | ESR1 | 873 | 0.092552771 | 0.008634595 | 10.71883176 | 11 | 2.26776808 | 1 | 0.293782445 | 0.11648139 | 0.000999001 |
| LRP | ESR1 | 873 | 0.003983811 | 0.001378483 | 2.889995666 | 7 | 2.26776808 | 1 | 0.387526415 | 0.078519851 | 0.011988012 |
| Saliency | ESR1 | 873 | 0.003559258 | 0.001265241 | 2.81310652 | 6 | 2.26776808 | 7.62E-23 | 0.603353246 | 0.370571551 | 0.000999001 |
| GNNExplainer | MYC | 2024 | 0.007318028 | 0.036939927 | 0.198106184 | 9 | 5.257689111 | 1 | 0.349572818 | -0.459645869 | 1 |
| IntegratedGradients | MYC | 2024 | 0.071533712 | 0.006394963 | 11.18594565 | 24 | 5.257689111 | 1 | 0.259422601 | 0.117066176 | 0.000999001 |
| LRP | MYC | 2024 | 0.003250373 | 0.001316379 | 2.469176116 | 9 | 5.257689111 | 1 | 0.374055419 | 0.059055006 | 0.008991009 |
| Saliency | MYC | 2024 | 0.002738158 | 0.001226657 | 2.232211373 | 9 | 5.257689111 | 9.02E-18 | 0.562456154 | 0.249469629 | 0.000999001 |
| GNNExplainer | TP53 | 2056 | 0.008113689 | 0.036848926 | 0.220187939 | 7 | 5.34081463 | 1 | 0.358377772 | -0.446461548 | 1 |
| IntegratedGradients | TP53 | 2056 | 0.043371479 | 0.009002752 | 4.817579971 | 19 | 5.34081463 | 1 | 0.268231582 | 0.096158235 | 0.000999001 |
| LRP | TP53 | 2056 | 0.002280938 | 0.001405956 | 1.62234006 | 9 | 5.34081463 | 1 | 0.383660848 | 0.03355381 | 0.098901099 |
| Saliency | TP53 | 2056 | 0.002699371 | 0.001231135 | 2.192586766 | 12 | 5.34081463 | 2.11E-32 | 0.586865435 | 0.240433509 | 0.000999001 |

**Supplementary Table ST3 Importance of Known Disease Hubs (Gene + 1 -hop Neighborhood) Per Explainer.**

Assessment of explainer attribution around canonical breast cancer genes BRCA1, ESR1, MYC, and TP53 and their 1-hop STRING neighborhoods. For each method, enrichment ratio, top-50 overlap, AUC, statistical significance, and effect size quantify the extent to which attribution is concentrated within known disease-associated hub regions.

| score_name | seed_protein | seed_gene_name | observed_hub_score | hub_rank | percentile |
|---|---|---|---|---|---|
| IntegratedGradients | ENSP00000269305 | TP53 | 50.2439 | 79 | 0.9929 |
| Saliency | ENSP00000269305 | TP53 | 2.2622 | 1176 | 0.893 |
| LRP | ENSP00000269305 | TP53 | 3.3922 | 1253 | 0.886 |
| GNNExplainer | ENSP00000269305 | TP53 | 0.234 | 8611 | 0.2162 |
| Consensus (Untuned) | ENSP00000269305 | TP53 | 982.849 | 5 | 0.9996 |
| Consensus (Tuned) | ENSP00000269305 | TP53 | 1071.0542 | 5 | 0.9996 |
| IntegratedGradients | ENSP00000418960 | BRCA1 | 24.9973 | 900 | 0.9182 |
| Saliency | ENSP00000418960 | BRCA1 | 1.7651 | 3007 | 0.7264 |
| LRP | ENSP00000418960 | BRCA1 | 1.1363 | 5479 | 0.5013 |
| GNNExplainer | ENSP00000418960 | BRCA1 | 0.2255 | 8948 | 0.1855 |
| Consensus (Untuned) | ENSP00000418960 | BRCA1 | 27.6991 | 149 | 0.9865 |
| Consensus (Tuned) | ENSP00000418960 | BRCA1 | 70.2922 | 72 | 0.9935 |
| IntegratedGradients | ENSP00000405330 | ESR1 | 70.4163 | 20 | 0.9983 |
| Saliency | ENSP00000405330 | ESR1 | 1.7817 | 2899 | 0.7362 |
| LRP | ENSP00000405330 | ESR1 | 3.5509 | 1136 | 0.8967 |
| GNNExplainer | ENSP00000405330 | ESR1 | 0.2827 | 6714 | 0.3889 |
| Consensus (Untuned) | ENSP00000405330 | ESR1 | 59.419 | 70 | 0.9937 |
| Consensus (Tuned) | ENSP00000405330 | ESR1 | 75.8708 | 70 | 0.9937 |
| IntegratedGradients | ENSP00000478887 | MYC | 63.8746 | 27 | 0.9976 |
| Saliency | ENSP00000478887 | MYC | 1.5747 | 4263 | 0.612 |
| LRP | ENSP00000478887 | MYC | 1.7832 | 3291 | 0.7005 |
| GNNExplainer | ENSP00000478887 | MYC | 0.2561 | 7718 | 0.2975 |
| Consensus (Untuned) | ENSP00000478887 | MYC | 27.6422 | 151 | 0.9863 |
| Consensus (Tuned) | ENSP00000478887 | MYC | 93.8248 | 54 | 0.9952 |

**Supplementary Table ST4 Candidate Hub Scores for Known Breast Cancer Hubs.** Ranking of TP53, BRCA1, ESR1, and MYC using shell-based hub scores derived from individual explainers and consensus scoring approaches. Observed hub scores, genome-wide ranks, and percentiles show how effectively each method prioritizes known breast cancer-associated hubs.

| score_name | R2_degree_on_hub_score | beta_intercept | beta_log_degree |
|---|---|---|---|
| IntegratedGradients | 0.196146627 | -10.73811187 | 4.58480383 |
| GNNExplainer | 0.084501642 | 0.841951327 | -0.099302629 |
| Consensus (Tuned) | 0.023033314 | -17.2287063 | 5.527803919 |
| Consensus (Untuned) | 0.022663342 | -14.06612874 | 4.637237928 |
| LRP | 0.000853295 | 1.565110474 | 0.062698384 |
| Saliency | 4.08E-06 | 1.555884918 | -0.001451844 |

**Supplementary Table ST5 HubScore Degree $R^2$ Table.**

Linear regression analysis of hub scores against log-transformed node degree in the STRING network. The table reports $R^2$ values and regression coefficients for each explainer and consensus method, quantifying the extent to which hub prioritization is explained by network degree alone.

| score_name | seed_gene_name | raw_hub_score | raw_rank | raw_percentile | degree_resid_rank | degree_resid_percentile |
|---|---|---|---|---|---|---|
| IntegratedGradients | TP53 | 50.24388386 | 79 | 0.992899408 | 194 | 0.982430587 |
| IntegratedGradients | BRCA1 | 24.99734917 | 900 | 0.918161129 | 2037 | 0.81465635 |
| IntegratedGradients | ESR1 | 70.41629662 | 20 | 0.998270369 | 27 | 0.997633136 |
| IntegratedGradients | MYC | 63.87459513 | 27 | 0.997633136 | 51 | 0.995448339 |
| Saliency | TP53 | 2.262153822 | 1176 | 0.893035958 | 1167 | 0.893855257 |
| Saliency | BRCA1 | 1.765135462 | 3007 | 0.726354119 | 2986 | 0.728265817 |
| Saliency | ESR1 | 1.781719403 | 2899 | 0.736185708 | 2871 | 0.738734638 |
| Saliency | MYC | 1.574657644 | 4263 | 0.612016386 | 4227 | 0.615293582 |
| LRP | TP53 | 3.392238678 | 1253 | 0.8860264 | 1394 | 0.873190715 |
| LRP | BRCA1 | 1.136315205 | 5479 | 0.501319982 | 6555 | 0.403368229 |
| LRP | ESR1 | 3.550931882 | 1136 | 0.896677287 | 1235 | 0.887664998 |
| LRP | MYC | 1.783204259 | 3291 | 0.700500683 | 3778 | 0.656167501 |
| GNNExplainer | TP53 | 0.233993442 | 8611 | 0.216203914 | 1865 | 0.830314065 |
| GNNExplainer | BRCA1 | 0.225546434 | 8948 | 0.185525717 | 3252 | 0.704050979 |
| GNNExplainer | ESR1 | 0.282707421 | 6714 | 0.388893946 | 2210 | 0.798907601 |
| GNNExplainer | MYC | 0.256149844 | 7718 | 0.297496586 | 1732 | 0.842421484 |
| Consensus (Untuned) | TP53 | 982.8489567 | 5 | 0.999633968 | 5 | 0.999633968 |
| Consensus (Untuned) | BRCA1 | 27.69912398 | 149 | 0.986456808 | 296 | 0.973005124 |
| Consensus (Untuned) | ESR1 | 59.41898088 | 70 | 0.993685944 | 75 | 0.993228404 |
| Consensus (Untuned) | MYC | 27.64216506 | 151 | 0.986273792 | 869 | 0.92057101 |
| Consensus (Tuned) | TP53 | 1071.054197 | 5 | 0.999633968 | 5 | 0.999633968 |
| Consensus (Tuned) | BRCA1 | 70.29218604 | 72 | 0.993502928 | 76 | 0.993136896 |
| Consensus (Tuned) | ESR1 | 75.87078469 | 70 | 0.993685944 | 70 | 0.993685944 |
| Consensus (Tuned) | MYC | 93.82481325 | 54 | 0.995150073 | 58 | 0.994784041 |

**Supplementary Table ST6 HubScore Raw vs Degree Deresidualized.**

Comparison of hub score rankings before and after removing the linear contribution of node degree. Raw and residualized ranks and percentiles are shown for TP53, BRCA1, ESR1, and MYC across individual explainers and consensus methods to assess robustness to degree confounding.

**Supplementary Table ST7 Reactome pathway enrichment of top candidate hubs. (Attached as supplementary file.)**

Reactome enrichment analysis of the top 100 candidate hub genes identified by each explainer-derived scoring method. Pathways are ranked by adjusted p-value and reported with enrichment statistics, overlap, and within-method rank to compare biological programs recovered by individual and consensus explanation strategies.

| Score Name | Term | Adjusted P-value | P-value | Odds Ratio | Comb. Score | Overlap | Rank Within Method |
|---|---|---|---|---|---|---|---|
| IntegratedGradients | Peptide Chain Elongation R-HSA-156902 | 8.31E-24 | 1.30E-26 | 69.4356 | 4138.5233 | 19/86 | 1 |
| IntegratedGradients | Eukaryotic Translation Elongation R-HSA-156842 | 1.08E-23 | 3.37E-26 | 65.5105 | 3842.2474 | 19/90 | 2 |
| IntegratedGradients | Selenocysteine Synthesis R-HSA-2408557 | 1.62E-20 | 1.27E-22 | 55.6295 | 2804.6175 | 17/90 | 3 |
| IntegratedGradients | Eukaryotic Translation Termination R-HSA-72764 | 1.62E-20 | 1.27E-22 | 55.6295 | 2804.6175 | 17/90 | 4 |
| IntegratedGradients | Viral mRNA Translation R-HSA-192823 | 1.62E-20 | 1.27E-22 | 55.6295 | 2804.6175 | 17/90 | 5 |
| IntegratedGradients | Nonsense Mediated Decay (NMD) Independent Of Exon Junction Complex (EJC) R-HSA-975956 | 2.02E-20 | 1.90E-22 | 54.1406 | 2707.7211 | 17/92 | 6 |
| IntegratedGradients | Response Of EIF2AK4 (GCN2) To Amino Acid Deficiency R-HSA-9633012 | 4.81E-20 | 6.03E-22 | 50.115 | 2448.6061 | 17/98 | 7 |
| IntegratedGradients | Formation Of A Pool Of Free 40S Subunits R-HSA-72689 | 4.81E-20 | 6.03E-22 | 50.115 | 2448.6061 | 17/98 | 8 |
| IntegratedGradients | SARS-CoV-2 Modulates Host Translation Machinery R-HSA-9754678 | 5.00E-20 | 7.06E-22 | 98.0049 | 4773.1154 | 14/47 | 9 |
| IntegratedGradients | Formation Of Ternary Complex, And Subsequently, 43S Complex R-HSA-72695 | 1.22E-19 | 1.91E-21 | 89.8243 | 4285.0704 | 14/50 | 10 |
| IntegratedGradients | Influenza Infection R-HSA-168255 | 1.47E-19 | 2.53E-21 | 33.5908 | 1593.1022 | 19/157 | 11 |
| IntegratedGradients | L13a-mediated Translational Silencing Of Ceruloplasmin Expression R-HSA-156827 | 1.71E-19 | 3.49E-21 | 44.5853 | 2100.1105 | 17/108 | 12 |
| IntegratedGradients | SRP-dependent Cotranslational Protein Targeting To Membrane R-HSA-1799339 | 1.71E-19 | 3.49E-21 | 44.5853 | 2100.1105 | 17/108 | 13 |
| IntegratedGradients | GTP Hydrolysis And Joining Of 60S Ribosomal Subunit R-HSA-72706 | 1.88E-19 | 4.12E-21 | 44.0985 | 2069.8746 | 17/109 | 14 |
| IntegratedGradients | Influenza Viral RNA Transcription And Replication R-HSA-168273 | 2.68E-19 | 6.65E-21 | 36.4888 | 1695.2358 | 18/137 | 15 |
| IntegratedGradients | Nonsense Mediated Decay (NMD) Enhanced By Exon Junction Complex (EJC) R-HSA-975957 | 2.68E-19 | 6.71E-21 | 42.6994 | 1983.4149 | 17/112 | 16 |
| IntegratedGradients | Regulation Of Expression Of SLITs And ROBOs R-HSA-9010553 | 3.17E-19 | 8.44E-21 | 31.3053 | 1446.9815 | 19/167 | 17 |
| IntegratedGradients | Selenoamino Acid Metabolism R-HSA-2408522 | 3.26E-19 | 9.21E-21 | 41.8148 | 1929.0828 | 17/114 | 18 |
| IntegratedGradients | Cap-dependent Translation Initiation R-HSA-72737 | 4.22E-19 | 1.26E-20 | 40.9659 | 1877.1981 | 17/116 | 19 |
| IntegratedGradients | Ribosomal Scanning And Start Codon Recognition R-HSA-72702 | 4.64E-19 | 1.53E-20 | 75.1752 | 3430.1288 | 14/57 | 20 |
| Saliency | GABA Receptor Activation R-HSA-977443 | 6.11E-26 | 4.16E-28 | 106.3242 | 6703.4855 | 18/59 | 1 |
| Saliency | Neurotransmitter Receptors And Postsynaptic Signal Transmission R-HSA-112314 | 9.18E-17 | 1.25E-18 | 26.4164 | 1088.9877 | 18/182 | 2 |
| Saliency | Transmission Across Chemical Synapses R-HSA-112315 | 1.36E-14 | 2.78E-16 | 18.9397 | 678.4036 | 18/246 | 3 |
| Saliency | Neuronal System R-HSA-112316 | 2.47E-11 | 6.73E-13 | 11.6508 | 326.5308 | 18/386 | 4 |
| Saliency | Metabolism Of Nucleotides R-HSA-15869 | 1.29E-09 | 4.38E-11 | 25.902 | 617.7849 | Oct-95 | 5 |
| Saliency | Digestion Of Dietary Lipid R-HSA-192456 | 1.44E-09 | 5.88E-11 | 523.6316 | 12334.8213 | 7-May | 6 |
| Saliency | Signaling By ERBB4 R-HSA-1236394 | 7.29E-09 | 3.97E-10 | 35.228 | 762.6148 | Aug-57 | 7 |
| Saliency | Nucleotide Catabolism R-HSA-8956319 | 7.29E-09 | 3.78E-10 | 53.4194 | 1158.9326 | Jul-35 | 8 |
| Saliency | Pyrimidine Catabolism R-HSA-73621 | 3.55E-08 | 2.18E-09 | 149.5714 | 2983.3539 | 12-May | 9 |
| Saliency | Purine Catabolism R-HSA-74259 | 2.45E-07 | 1.67E-08 | 87.2281 | 1562.2543 | 17-May | 10 |
| Saliency | Digestion R-HSA-8935690 | 1.18E-06 | 8.85E-08 | 58.1345 | 944.1262 | 23-May | 11 |
| Saliency | Digestion And Absorption R-HSA-8963743 | 3.10E-06 | 2.53E-07 | 45.4851 | 690.8431 | 28-May | 12 |
| Saliency | Nucleotide Salvage R-HSA-8956321 | 0.002258489 | 0.0001997 | 30.7423 | 261.8793 | 23-Mar | 13 |
| Saliency | Pyrimidine Salvage R-HSA-73614 | 0.013879321 | 0.0013218 | 45.1043 | 298.9843 | 11-Feb | 14 |
| Saliency | Purine Salvage R-HSA-74217 | 0.018251795 | 0.0018624 | 36.8998 | 231.9476 | 13-Feb | 15 |
| Saliency | Hyaluronan Metabolism R-HSA-2142845 | 0.02606875 | 0.0028374 | 28.9883 | 170.0126 | 16-Feb | 16 |
| Saliency | Signaling By Receptor Tyrosine Kinases R-HSA-9006934 | 0.029677827 | 0.0034321 | 3.459 | 19.6284 | 8/496 | 17 |
| Saliency | Nitric Oxide Stimulates Guanylate Cyclase R-HSA-392154 | 0.039897329 | 0.0048854 | 21.3545 | 113.6379 | 21-Feb | 18 |
| Saliency | Metabolism R-HSA-1430728 | 0.044976713 | 0.0058133 | 2.0649 | 10.6292 | 19/2049 | 19 |
| Saliency | ROS And RNS Production In Phagocytes R-HSA-1222556 | 0.102612727 | 0.0139609 | 11.9244 | 50.9349 | Feb-36 | 20 |
| LRP | Cytochrome P450 - Arranged By Substrate Type R-HSA-211897 | 6.66E-06 | 3.47E-08 | 25.7497 | 442.3104 | Jul-65 | 1 |
| LRP | Phase I - Functionalization Of Compounds R-HSA-211945 | 8.67E-05 | 9.03E-07 | 15.3665 | 213.8683 | 7/104 | 2 |
| LRP | Biosynthesis Of Maresin-Like SPMs R-HSA-9027307 | 0.000153555 | 2.40E-06 | 205.1237 | 2654.3695 | 6-Mar | 3 |
| LRP | Biosynthesis Of Maresins R-HSA-9018682 | 0.000256103 | 6.67E-06 | 123.0619 | 1466.6497 | 8-Mar | 4 |
| LRP | CREB3 Factors Activate Genes R-HSA-8874211 | 0.000256103 | 6.67E-06 | 123.0619 | 1466.6497 | 8-Mar | 5 |
| LRP | Synthesis Of (16-20)-Hydroxyeicosatetraenoic Acids (HETE) R-HSA-2142816 | 0.000318967 | 9.97E-06 | 102.5464 | 1180.9406 | 9-Mar | 6 |
| LRP | Biological Oxidations R-HSA-211859 | 0.000378462 | 1.38E-05 | 8.1532 | 91.2424 | 8/218 | 7 |
| LRP | Biosynthesis Of DHA-derived SPMs R-HSA-9018677 | 0.001881079 | 7.84E-05 | 43.9308 | 415.32 | 17-Mar | 8 |
| LRP | Biosynthesis Of Specialized Proresolving Mediators (SPMs) R-HSA-9018678 | 0.002365454 | 0.0001109 | 38.4356 | 350.0349 | 19-Mar | 9 |
| LRP | Xenobiotics R-HSA-211981 | 0.004366779 | 0.0002274 | 29.2769 | 245.5932 | 24-Mar | 10 |
| LRP | Endogenous Sterols R-HSA-211976 | 0.005674898 | 0.0003251 | 25.6134 | 205.709 | 27-Mar | 11 |
| LRP | RHOB GTPase Cycle R-HSA-9013026 | 0.006343244 | 0.0003965 | 12.7147 | 99.594 | Apr-69 | 12 |
| LRP | Synthesis Of Epoxy (EET) And Dihydroxyeicosatrienoic Acids (DHET) R-HSA-2142670 | 0.009319614 | 0.0006796 | 67.6667 | 493.5655 | 8-Feb | 13 |
| LRP | Relaxin Receptors R-HSA-444821 | 0.009319614 | 0.0006796 | 67.6667 | 493.5655 | 8-Feb | 14 |
| LRP | RHO GTPases Activate Rhotekin And Rhophilins R-HSA-5666185 | 0.0111471 | 0.0008709 | 57.9971 | 408.6487 | 9-Feb | 15 |
| LRP | Metabolism Of Lipids R-HSA-556833 | 0.012905586 | 0.0010755 | 3.2877 | 22.4715 | 11/732 | 16 |
| LRP | Synthesis Of Bile Acids And Bile Salts Via 27-Hydroxycholesterol R-HSA-193807 | 0.028131776 | 0.0024908 | 31.2198 | 187.1669 | 15-Feb | 17 |
| LRP | Arachidonic Acid Metabolism R-HSA-2142753 | 0.034334314 | 0.0032188 | 10.9595 | 62.8936 | Mar-59 | 18 |
| LRP | Aflatoxin Activation And Detoxification R-HSA-5423646 | 0.040461935 | 0.004004 | 23.8691 | 131.7684 | 19-Feb | 19 |
| LRP | Synaptic Adhesion-Like Molecules R-HSA-8849932 | 0.046899718 | 0.0048854 | 21.3545 | 113.6379 | 21-Feb | 20 |
| GNNExplainer | Expression And Translocation Of Olfactory Receptors R-HSA-9752946 | 7.11E-12 | 7.52E-14 | 12.2465 | 370.0787 | 19/393 | 1 |
| GNNExplainer | Olfactory Signaling Pathway R-HSA-381753 | 7.11E-12 | 1.08E-13 | 11.9851 | 357.8587 | 19/401 | 2 |
| GNNExplainer | Sensory Perception R-HSA-9709957 | 1.32E-11 | 3.00E-13 | 9.1671 | 264.3486 | 22/616 | 3 |
| GNNExplainer | Peptide Ligand-Binding Receptors R-HSA-375276 | 1.73E-06 | 5.23E-08 | 11.7766 | 197.4491 | 10/196 | 4 |
| GNNExplainer | Class A/1 (Rhodopsin-like Receptors) R-HSA-373076 | 0.000147995 | 5.61E-06 | 6.864 | 82.9974 | 10/327 | 5 |
| GNNExplainer | GPCR Ligand Binding R-HSA-500792 | 0.001865482 | 9.89E-05 | 4.8244 | 44.4865 | 10/458 | 6 |
| GNNExplainer | G Alpha (Q) Signaling Events R-HSA-416476 | 0.001865482 | 9.47E-05 | 7.2313 | 66.9985 | 7/212 | 7 |
| GNNExplainer | Sensory Perception Of Salty Taste R-HSA-9730628 | 0.006046131 | 0.0003664 | 101.5102 | 803.118 | 6-Feb | 8 |
| GNNExplainer | Orexin And Neuropeptides FF And QRFP Bind To Their Respective Receptors R-HSA-389397 | 0.008970128 | 0.0006796 | 67.6667 | 493.5655 | 8-Feb | 9 |
| GNNExplainer | Signaling By GPCR R-HSA-372790 | 0.008970128 | 0.0006564 | 3.5041 | 25.6801 | 11/689 | 10 |
| GNNExplainer | GPCR Downstream Signaling R-HSA-388396 | 0.012782642 | 0.0010652 | 3.5196 | 24.0903 | 10/619 | 11 |
| GNNExplainer | Lewis Blood Group Biosynthesis R-HSA-9037629 | 0.039536608 | 0.0035942 | 25.3622 | 142.7495 | 18-Feb | 12 |
| GNNExplainer | Phase 4 - Resting Membrane Potential R-HSA-5576886 | 0.040656463 | 0.004004 | 23.8691 | 131.7684 | 19-Feb | 13 |
| GNNExplainer | Blood Group Systems Biosynthesis R-HSA-9033658 | 0.050504114 | 0.0053565 | 20.2857 | 106.083 | 22-Feb | 14 |
| GNNExplainer | Synthesis Of PC R-HSA-1483191 | 0.070471337 | 0.0080081 | 16.2245 | 78.3205 | 27-Feb | 15 |
| GNNExplainer | Potassium Channels R-HSA-1296071 | 0.120220473 | 0.0145722 | 6.1859 | 26.1579 | 3/102 | 16 |
| GNNExplainer | Sensory Perception Of Sweet, Bitter, And Umami (Glutamate) Taste R-HSA-9717207 | 0.138836511 | 0.0178805 | 10.393 | 41.8219 | Feb-41 | 17 |
| GNNExplainer | Sensory Perception Of Taste R-HSA-9717189 | 0.169550756 | 0.0231206 | 9.0045 | 33.9204 | Feb-47 | 18 |
| GNNExplainer | Tachykinin Receptors Bind Tachykinins R-HSA-380095 | 0.171971884 | 0.0247535 | 50.2424 | 185.836 | 5-Jan | 19 |
| GNNExplainer | CREB3 Factors Activate Genes R-HSA-8874211 | 0.192199354 | 0.0393135 | 28.7056 | 92.8968 | 8-Jan | 20 |
| Consensus (Untuned) | Signaling By Interleukins R-HSA-449147 | 7.51E-19 | 7.48E-22 | 16.9078 | 822.4661 | 27/453 | 1 |
| Consensus (Untuned) | Immune System R-HSA-168256 | 3.07E-18 | 6.12E-21 | 8.0843 | 376.2593 | 46/1943 | 2 |
| Consensus (Untuned) | Disease R-HSA-1643685 | 1.03E-16 | 3.09E-19 | 7.7826 | 331.6905 | 42/1736 | 3 |
| Consensus (Untuned) | Diseases Of Signal Transduction By Growth Factor Receptors And Second Messengers R-HSA-5663202 | 1.62E-16 | 6.46E-19 | 15.3947 | 644.7775 | 24/424 | 4 |
| Consensus (Untuned) | Cellular Responses To Stress R-HSA-2262752 | 1.87E-16 | 9.34E-19 | 11.3205 | 469.9745 | 29/722 | 5 |
| Consensus (Untuned) | Cellular Responses To Stimuli R-HSA-8953897 | 2.62E-16 | 1.57E-18 | 11.0883 | 454.5862 | 29/736 | 6 |
| Consensus (Untuned) | Cytokine Signaling In Immune System R-HSA-1280215 | 7.22E-16 | 5.04E-18 | 11.0931 | 441.8329 | 28/702 | 7 |
| Consensus (Untuned) | Infectious Disease R-HSA-5663205 | 2.93E-15 | 2.33E-17 | 9.1643 | 350.956 | 31/961 | 8 |
| Consensus (Untuned) | Cellular Response To Heat Stress R-HSA-3371556 | 7.10E-15 | 6.37E-17 | 37.9494 | 1415.2314 | 14/99 | 9 |
| Consensus (Untuned) | Signal Transduction R-HSA-162582 | 5.09E-14 | 5.07E-16 | 5.9098 | 208.1314 | 45/2465 | 10 |
| Consensus (Untuned) | Signaling By ERBB2 R-HSA-1227986 | 6.29E-14 | 6.90E-16 | 64.6014 | 2255.2274 | Nov-49 | 11 |
| Consensus (Untuned) | Spry Regulation Of FGF Signaling R-HSA-1295596 | 2.83E-13 | 3.65E-15 | 216.2174 | 7187.7313 | 16-Aug | 12 |
| Consensus (Untuned) | Signaling By Receptor Tyrosine Kinases R-HSA-9006934 | 2.83E-13 | 3.67E-15 | 11.5593 | 384.2069 | 22/496 | 13 |
| Consensus (Untuned) | Vesicle-mediated Transport R-HSA-5653656 | 4.57E-13 | 6.38E-15 | 9.9358 | 324.7547 | 24/637 | 14 |
| Consensus (Untuned) | MAPK Family Signaling Cascades R-HSA-5683057 | 1.65E-12 | 2.46E-14 | 14.3415 | 449.3784 | 18/318 | 15 |
| Consensus (Untuned) | Chaperone Mediated Autophagy R-HSA-9613829 | 2.21E-12 | 3.52E-14 | 144.1159 | 4464.4607 | 20-Aug | 16 |
| Consensus (Untuned) | Platelet Activation, Signaling And Aggregation R-HSA-76002 | 8.76E-12 | 1.48E-13 | 15.7359 | 464.8193 | 16/254 | 17 |
| Consensus (Untuned) | Innate Immune System R-HSA-168249 | 3.83E-11 | 6.87E-13 | 6.932 | 194.1423 | 27/1035 | 18 |
| Consensus (Untuned) | SARS-CoV-2 Infection R-HSA-9694516 | 4.14E-11 | 7.84E-13 | 14.0061 | 390.4118 | 16/283 | 19 |
| Consensus (Untuned) | Downregulation Of ERBB2 Signaling R-HSA-8863795 | 5.27E-11 | 1.16E-12 | 82.3147 | 2262.547 | 29-Aug | 20 |
| Consensus (Tuned) | Immune System R-HSA-168256 | 7.31E-19 | 7.37E-22 | 8.4208 | 409.7471 | 47/1943 | 1 |
| Consensus (Tuned) | Signaling By Interleukins R-HSA-449147 | 6.19E-18 | 1.25E-20 | 16.0231 | 734.3239 | 26/453 | 2 |
| Consensus (Tuned) | Cellular Responses To Stress R-HSA-2262752 | 3.08E-16 | 9.34E-19 | 11.3205 | 469.9745 | 29/722 | 3 |
| Consensus (Tuned) | Cellular Responses To Stimuli R-HSA-8953897 | 3.88E-16 | 1.57E-18 | 11.0883 | 454.5862 | 29/736 | 4 |
| Consensus (Tuned) | Disease R-HSA-1643685 | 4.72E-16 | 2.38E-18 | 7.4637 | 302.8719 | 41/1736 | 5 |
| Consensus (Tuned) | Cytokine Signaling In Immune System R-HSA-1280215 | 8.32E-16 | 5.04E-18 | 11.0931 | 441.8329 | 28/702 | 6 |
| Consensus (Tuned) | Cellular Response To Heat Stress R-HSA-3371556 | 9.02E-15 | 6.37E-17 | 37.9494 | 1415.2314 | 14/99 | 7 |
| Consensus (Tuned) | Diseases Of Signal Transduction By Growth Factor Receptors And Second Messengers R-HSA-5663202 | 1.75E-14 | 1.41E-16 | 13.6802 | 499.2702 | 22/424 | 8 |
| Consensus (Tuned) | Infectious Disease R-HSA-5663205 | 2.33E-14 | 2.12E-16 | 8.7321 | 315.1464 | 30/961 | 9 |
| Consensus (Tuned) | Signal Transduction R-HSA-162582 | 5.03E-14 | 5.07E-16 | 5.9098 | 208.1314 | 45/2465 | 10 |
| Consensus (Tuned) | Signaling By Receptor Tyrosine Kinases R-HSA-9006934 | 3.31E-13 | 3.67E-15 | 11.5593 | 384.2069 | 22/496 | 11 |
| Consensus (Tuned) | Downregulation Of ERBB2 Signaling R-HSA-8863795 | 1.03E-12 | 1.25E-14 | 98.3077 | 3147.4893 | 29-Sep | 12 |
| Consensus (Tuned) | Chaperone Mediated Autophagy R-HSA-9613829 | 2.68E-12 | 3.52E-14 | 144.1159 | 4464.4607 | 20-Aug | 13 |
| Consensus (Tuned) | Signaling By ERBB2 R-HSA-1227986 | 3.05E-12 | 4.30E-14 | 56.584 | 1741.4736 | Oct-49 | 14 |
| Consensus (Tuned) | Signaling By Non-Receptor Tyrosine Kinases R-HSA-9006927 | 5.41E-12 | 8.18E-14 | 52.5344 | 1583.0755 | Oct-52 | 15 |
| Consensus (Tuned) | Signaling By WNT R-HSA-195721 | 6.05E-12 | 9.77E-14 | 14.5096 | 434.6562 | 17/294 | 16 |
| Consensus (Tuned) | Downregulation Of ERBB2:ERBB3 Signaling R-HSA-1358803 | 6.16E-12 | 1.06E-13 | 249.5663 | 7456.7965 | 13-Jul | 17 |
| Consensus (Tuned) | Ovarian Tumor Domain Proteases R-HSA-5689896 | 1.39E-11 | 2.53E-13 | 65.5055 | 1900.025 | Sep-39 | 18 |
| Consensus (Tuned) | Innate Immune System R-HSA-168249 | 3.58E-11 | 6.87E-13 | 6.932 | 194.1423 | 27/1035 | 19 |
| Consensus (Tuned) | SARS-CoV-2 Infection R-HSA-9694516 | 3.88E-11 | 7.84E-13 | 14.0061 | 390.4118 | 16/283 | 20 |